\documentclass[10pt,journal,compsoc]{IEEEtran}
\newif\ifpeerreview

\peerreviewfalse

\usepackage[nocompress]{cite}
\usepackage{url}
\usepackage{amsmath,amssymb,graphicx}

\usepackage{bm}
\DeclareMathOperator*{\argmin}{arg\,min}

\usepackage[switch]{lineno}

\newcommand{\paperID}{38}

\title{Dynamic Structured Illumination Microscopy \\ with a Neural Space-time Model}

\author{Ruiming~Cao, Fanglin~Linda~Liu, Li-Hao~Yeh,
        and~Laura~Waller% <-this % stops a space
\IEEEcompsocitemizethanks{\IEEEcompsocthanksitem R. Cao is with the Department of Bioengineering, F.L. Liu and L. Waller are with the Department
of Electrical Engineering and Computer Sciences, University of California, Berkeley, Berkeley,
CA, 94720. L.-H. Yeh is with the Chan Zuckerberg Biohub, San Francisco, CA, 94158.\protect\\
E-mail: rcao@berkeley.edu.}% <-this % stops an unwanted space
}

\begin{document}

\IEEEtitleabstractindextext{%
\begin{abstract}
Structured illumination microscopy (SIM) reconstructs a super-resolved image from multiple raw images captured with different illumination patterns; hence, acquisition speed is limited, making it unsuitable for dynamic scenes. We propose a new method, Speckle Flow SIM, that uses static patterned illumination with moving samples and models the sample motion during data capture in order to reconstruct the dynamic scene with super-resolution. Speckle Flow SIM relies on sample motion to capture a sequence of raw images. The spatio-temporal relationship of the dynamic scene is modeled using a neural space-time model with coordinate-based multi-layer perceptrons (MLPs), and the motion dynamics and the super-resolved scene are jointly recovered.  We validate Speckle Flow SIM for coherent imaging in simulation and build a simple, inexpensive experimental setup with off-the-shelf components. We demonstrate that Speckle Flow SIM can reconstruct a dynamic scene with deformable motion and 1.88$\times$ the diffraction-limited resolution in experiment.
\end{abstract}

}

% Make Title
\ifpeerreview
\linenumbers \linenumbersep 15pt\relax 
\author{Paper ID \paperID\IEEEcompsocitemizethanks{\IEEEcompsocthanksitem This paper is under review for ICCP 2022 and the PAMI special issue on computational photography. Do not distribute.}}
\markboth{Anonymous ICCP 2022 submission ID \paperID}%
{}
\fi
\maketitle
\thispagestyle{empty}
\IEEEdisplaynontitleabstractindextext

\IEEEraisesectionheading{
  \section{Introduction}\label{sec:introduction}
}
% super res imaging, general, usage/importance
\IEEEPARstart{T}{he} spatial resolution of any optical imaging system is fundamentally limited by diffraction, which depends on the system's numerical aperture (NA) and the wavelength of the light. Super-resolution microscopy has enabled the observation of sub-cellular and molecular structures beyond the diffraction limit and was recognized by the 2014 Nobel prize in chemistry for its fundamental role in today's biology research~\cite{mockl2014super,wu2018faster}.

% sim, speckle sim, pros and cons, ways to get diverse measurements
Structured illumination microscopy (SIM) is a practical super-resolution method that uses patterned illumination to encode high-frequency information from Moiré patterns, and then computationally decodes the image, enabling 2$\times$ better resolution than the diffraction limit~\cite{gustafsson2000surpassing}. Compared with other super-resolution methods (e.g., STED~\cite{hell1994breaking}, PALM~\cite{betzig2006imaging}), SIM has faster frame rates, lower photo-toxicity~\cite{godin2014super}, and is compatible with both brightfield~\cite{chowdhury2013structured} and generic fluorescence methods~\cite{gustafsson2000surpassing}. Although SIM usually uses sinusoidal illumination, it can also be implemented with random unknown speckle illumination, called speckle SIM~\cite{mudry2012structured}. The speckle is generated by a plane wave passing through a scattering layer and thus is easy to set up in experiment and does not require a calibration step for the structured pattern~\cite{wicker2013non, lal2016structured}.

In both sinusoidal and speckle SIM, multiple images must be captured to achieve super-resolution; hence, SIM trades off the system's temporal resolution for spatial super-resolution. Sinusoidal SIM methods usually take $\sim$10 diffraction-limited raw measurements, each with a different shift or rotation of the sinusoidal illumination pattern, to reconstruct a 2$\times$ super-resolved image~\cite{gustafsson2000surpassing}. Speckle SIM methods often require dozens of randomized speckle illuminations~\cite{mudry2012structured} or shifted speckle patterns~\cite{yeh2019computational, yeh2019speckle}. In both cases, the multi-shot nature of SIM effectively reduces the frame rate by at least $\sim$10-fold. If the scene contains motion during the multi-shot acquisition, the recovered super-resolved image will suffer from motion artifacts~\cite{forster2016motion}. Here, we explore methods for modeling the spatio-temporal relationship of the dynamic scene in order to reconstruct images without motion artifacts.

% compressive rep, neural representation, coordinate-based network, space-time using coord-based nn
A dynamic, super-resolved scene can be modeled using neural representations, which are a class of methods that encode high-dimensional information into an untrained deep neural network and store the information compressively in the neural network's weights~\cite{lucas2018using}. A coordinate-based neural network is a type of neural representation which maps a matrix coordinate to its corresponding value on the matrix, usually via a multi-layer perceptron (MLP)~\cite{stanley2007compositional}. Coordinate-based MLPs were first demonstrated to model for 3D scenes~\cite{mildenhall2020nerf} or 3D geometries~\cite{sitzmann2019scene, park2019deepsdf}. The spatio-temporal relationship can be similarly represented using coordinate-based MLPs by adding a time coordinate~\cite{park2021nerfies,pumarola2021d}.

\begin{figure*}[]
\centering
\includegraphics[width=\textwidth]{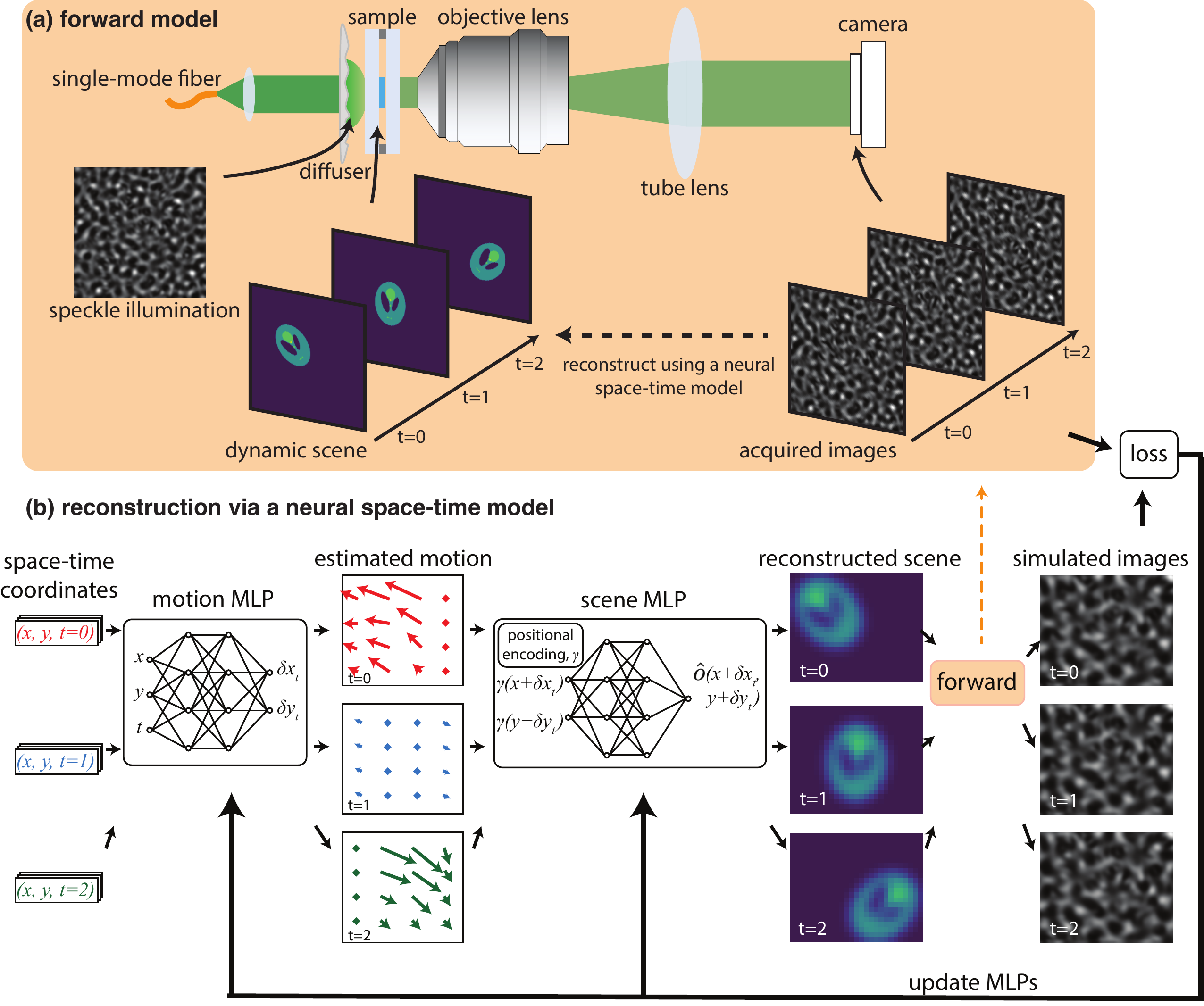}
\caption{Overview of Speckle Flow Structured Illumination Microscopy (SIM). (a) A plane wave passes through a thin scattering layer to generate speckle-structured illumination at the sample. The microscope, a 4$f$ system with an objective lens and a tube lens, then magnifies the image and the intensity is captured at the image plane by a CMOS sensor. The speckle illumination pattern is calibrated in advance and remains fixed while an image sequence of the dynamic scene is captured. The dynamic scene is modeled and simultaneously recovered with resolution beyond the diffraction limit. (b) The neural space-time model represents a dynamic scene using a motion multi-layer perceptron (MLP) and a scene MLP. The motion MLP takes a space-time coordinate, $\left( \bm{r}, t \right)=\left( x, y, t \right)$, corresponding to a pixel measured at a particular timepoint and estimates its displacement at $t$ relative to the time-independent scene stored in the scene MLP, $\delta \bm{r}_t = \left( \delta x_t, \delta y_t \right)$. The motion-accounted spatial coordinate, $\bm{r} + \delta \bm{r}_t$, is then fed into the scene MLP to query the corresponding value for the coordinate. This process is repeated for each coordinate to build up the entire scene replicated by the neural space-time model. During the reconstruction, the weights of the two MLPs are updated to minimize the difference (loss) between the acquired images and simulated images from the forward model.}
\label{fig:overview}
\end{figure*}

% our proposal: speckle sim for info compression, scene motion as a contrast mechanism, neural mlp as a novel high-dimensional rep for dynamic recon
In this work, we develop a new SIM method, called Speckle Flow SIM, that spatially super-resolves a dynamic scene using a neural space-time model. In Speckle Flow SIM, we modulate the scene with speckle illumination to encode high-frequency information into diffraction-limited images. Unlike previous speckle SIM systems that change the speckle to acquire sufficiently diversified information for a well-posed reconstruction, we maintain the speckle illumination unchanged but rely on the scene dynamics for the measurement diversity (see Fig.~\ref{fig:overview}(a)). We model the spatio-temporal relationship of the dynamic scene using a neural space-time model with coordinate-based MLPs~\cite{stanley2007compositional} and jointly recover the motion dynamics and the super-resolved scene. This allows Speckle Flow SIM to spatially super-resolve a dynamic scene with deformation motion. We validated Speckle Flow SIM for coherent imaging in simulation and experimentally demonstrated it using a simple, inexpensive setup.

\section{Related Work}
\label{sec:related}
% SIM, speckle SIM, li-hao's
Sinusoidal SIM is widely used to achieve up to two times better than diffraction-limited resolution~\cite{gustafsson2000surpassing}. Saturated SIM exploits the nonlinear response of saturated fluorescence and has unlimited theoretical resolution in the noise-free case and typically 5$\times$ super-resolution in experiments~\cite{gustafsson2005nonlinear}. SIM is compatible with both coherent~\cite{chowdhury2013structured} and fluorescence systems~\cite{gustafsson2000surpassing}, but saturated SIM only works for fluorescence systems. Speckle illumination is an alternative implementation of sinusoidal structured illumination, which also modulates high-frequency information into the diffraction-limited system~\cite{mudry2012structured, mangeat2021super, yeh2017structured,yeh2019computational, yeh2019speckle}. Unlike sinusoidal illumination, speckle illumination is random and requires either additional prior statistical information~\cite{mudry2012structured, yeh2017structured} or joint speckle calibration~\cite{yeh2019computational, yeh2019speckle} for a super-resolved image reconstruction. Speckle SIM has been experimentally demonstrated with a 1.6$\times$ resolution gain in a high-NA system~\cite{mudry2012structured} and a 5$\times$ resolution gain in a low-NA system~\cite{yeh2019computational}.

% temporal tradeoff, improve temporal resolution
SIM collects information beyond the diffraction limit by acquiring multiple raw images with varying illumination, which reduces the temporal resolution by an order-of-magnitude or more (depending on the number of raw images required for each super-resolved reconstruction). Previous study achieved video rate (10-20 Hz) SIM using a ferroelectric liquid crystal spatial light modulator (SLM) for a rapid update of structured light and a electron-multiplying charge coupled device (EMCCD) for a high-frame rate acquisition~\cite{kner2009super, lu2015fastsim}. \textit{Mangeat et al.} similarly demonstrated a high-frame rate version of speckle SIM using a SLM and a scientific CMOS camera~\cite{mangeat2021super}. While these methods improved the overall frame rate of the system, the scene is still assumed to be static for each of the raw images taken at different timepoints. To improve the SIM reconstruction by accounting for the motion of the scene, \textit{Shroff et al.} estimated the phase shift of the sinusoidal illumination to correct for a small global translational motion of the sample~\cite{shroff2010lateral}. \textit{Turcotte et al.} achieved dynamic in-vivo SIM imaging in the mouse brain by setting up a short exposure time and registering raw images to digitally remove motion artifacts~\cite{turcotte2019dynamic}. However, these studies assume a simple motion (e.g., rigid global motion) and do not aim to retain the motion dynamics as a final product of the reconstruction. In this study, we improve the temporal resolution by embedding the spatio-temporal relationship into the forward model to account for and recover the dynamics for each raw image, including non rigid-body motion. %Our method also does not require dedicated fast hardware.

% space-time model
Space-time model is a lasting theme in the computational imaging community. This modeling is especially useful when a multi-shot system is used to image a dynamic scene, such that the raw images acquired at different timepoints capture the scene at different states of the dynamics. A common strategy is to meter the motion and co-register acquired images for a static image reconstruction unaffected by the motion. For example, to improve the noise performance in low-light, smartphone cameras take a burst of frames and align them to compensate for motion during the image signal processing pipeline~\cite{hasinoff2016burst}. Motion differential phase contrast (DPC) microscopy used a navigation color channel to detect motion and wraps raw images into a single reference frame for the recovery of quantitative phase~\cite{kellman2018motion}. However, the raw images of speckle SIM do not provide enough information to estimate the motion without reconstructing the object. We instead opt to model the motion dynamics of the scene in a compressive way and simultaneously recover the scene and the motion in our reconstruction. \textit{Pnevmatikakis, et al.} expressed the fluorescence images of neural activity as a product of a spatial matrix and a temporal matrix to represent calcium imaging~\cite{pnevmatikakis2016simultaneous}. Instead of the matrix decomposition, we use neural representations to disentangle the spatio-temporal relationship.

\begin{figure*}[t]
\centering
\includegraphics[width=0.92\textwidth]{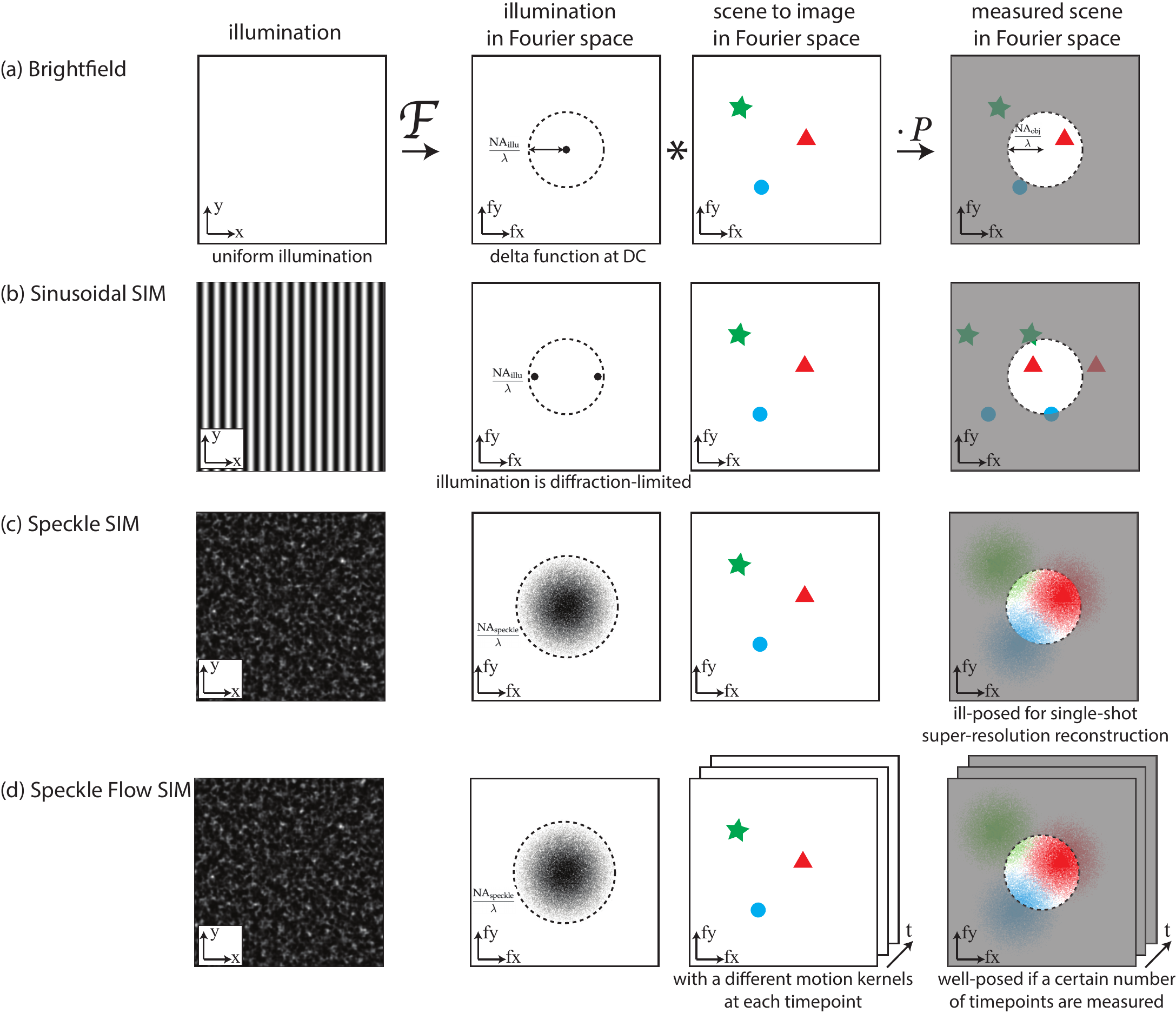}
\caption{Illustrations of spatial frequency information for (a) brightfield microscopy, (b) sinusoidal structured illumination microscopy (SIM), (c) speckle SIM, and (d) Speckle Flow SIM. The first column shows the illumination intensity at the sample. The second column is the amplitude of the illumination in Fourier space. The third column illustrates the spatial frequency information of a sample scene we want to image (note that this scene is sparse in Fourier space only for simplicity of visualization). The last column shows the measured spatial frequency bandwidth in Fourier space after passing through a diffraction-limited microscope (the grayed-out areas cannot be measured). More details in Sec.~\ref{sec:theory}.}
\label{fig:SIMintro}
\end{figure*}

% non-matrix rep, cnn rep, coordinate-based mlp rep
Neural representations use untrained artificial neural networks to reproduce a high-dimensional matrix or tensor in a compressive way for image reconstruction~\cite{lucas2018using}. The coordinate-based MLP~\cite{stanley2007compositional} has been used successfully in recent years in novel view synthesis for static~\cite{mildenhall2020nerf} and dynamic scenes\cite{park2021nerfies, pumarola2021d}, 3D geometry representation~\cite{sitzmann2019scene, sitzmann2020implicit, martel2021acorn, park2019deepsdf}, and partial differential equation solution~\cite{raissi2019physics}. Coordinate-based MLP maps a coordinate to its corresponding value, and it is suitable to represent a smooth, continuous object. Recent works improve the coordinate-based MLP's capacity for high-frequency information with sinusoidal Fourier features~\cite{mildenhall2020nerf, tancik2020fourier} or periodic activation functions~\cite{sitzmann2020implicit, martel2021acorn}. The coordinate-based MLP has also demonstrated an improved 3D reconstruction in the settings of computed tomography~\cite{sun2021coil} and optical diffraction tomography~\cite{liu2021zero}.

\section{Methods}
\label{sec:methods}

\subsection{Theory: Structured Illumination Microscopy}
\label{sec:theory}

% regular SIM for super-res, ok to use U instead of I here
In this section, we review the theory of sinusoidal and speckle structured illumination for super-resolution in a coherent imaging system. In a diffraction-limited system with the incident illumination field $U_{\text{in}}$ and the scene $o$, the measured output field, $U_{\text{out}}$, can be expressed as 
\begin{equation}
    U_{\text{out}}\left( \bm{r} \right) = \mathcal{F}^{-1} \left[ \mathcal{F} \left( U_{\text{in}}\left( \bm{r} \right) \cdot o\left( \bm{r} \right) \right) \cdot P\left( \bm{u} \right) \right],
\end{equation}
where $\mathcal{F}$ denotes 2D Fourier transform, and $\bm{r}$ and $\bm{u}$ are spatial coordinates in real and frequency space. The magnification is assumed to be 1 for simplicity. The pupil function, $P$, is a circular binary mask defined as $P\left(\bm{u}\right)=1 \text{ for } |\bm{u}| \leq \text{NA}_{\text{obj}} / \lambda, P\left(\bm{u}\right)=0$ \text{ if otherwise}, where $\lambda$ is the wavelength of the illumination and $\text{NA}_{\text{obj}}$ is the NA of the objective lens. In the case of brightfield illumination, $U_{\text{in}}$ is uniform and the pupil function directly low-pass filters the high-frequency information of the scene (Fig.~\ref{fig:SIMintro}(a)). The diffraction-limited resolution is set by the reciprocal of the pupil function bandwidth, i.e., $\lambda /  \text{NA}_{\text{obj}} $.

In sinusoidal SIM, the incident illumination is sinusoidal such that $U_{\text{in}} \left( \bm{r} \right)= \cos{\left(2\pi \bm{v_0} \cdot \bm{r} \right)}$, where $\bm{v_0}$ is the spatial frequency of the illumination pattern. This frequency shifts the spectrum of the scene such that higher spatial frequencies can be measured (Fig.~\ref{fig:SIMintro}(b)), enabling super-resolution. The raw measurement of SIM can be expressed as
\begin{equation}
    U_{\text{SIM}} \left( \bm{r} \right) = \mathcal{F}^{-1} \left[ \frac{\Tilde{o}\left( \bm{u} - \bm{v_0} \right) +  \Tilde{o}\left( \bm{u} + \bm{v_0} \right)}{2} \cdot P \left( \bm{u} \right) \right],
\end{equation}
where $\Tilde{\cdot}$ denotes the quantity in 2D Fourier transform space. The frequency of the sinusoidal pattern, $\bm{v_0}$, is also diffraction-limited in the far field such that $|\bm{v_0}| \leq \text{NA}_{\text{illu}} / \lambda$. The final resolution is  $\lambda / \left(  \text{NA}_{\text{obj}} +  \text{NA}_{\text{illu}} \right)$ which gives a 2$\times$ resolution gain when $\text{NA}_{\text{obj}} \approx \text{NA}_{\text{illu}}$. As each measurement $U_{\text{SIM}}$ is band-limited by the pupil function $P$ and only observes a fraction of the super-resolved scene $\Tilde{o}$, multiple measurements are needed to recover a single image.

% speckle as structured illumination
Speckle SIM uses random speckle illumination, $U_\text{sp}$, instead of sinusoidal illumination. The random speckle in Fourier transform space, $\Tilde{U}_{\text{sp}}$, also contains features from higher spatial frequencies, so super-resolved information can be encoded into a diffraction-limited measurement as in sinusoidal SIM (Fig.~\ref{fig:SIMintro}(c)). A speckle SIM measurement can be expressed mathematically as
\begin{equation}
\begin{split}
    U_{\text{SpeckleSIM}} \left( \bm{r} \right) &= \mathcal{F}^{-1} \left[ \mathcal{F} \left( U_{\text{sp}}\left( \bm{r} \right) \cdot o\left( \bm{r} \right) \right) \cdot P\left( \bm{u} \right) \right] \\ 
    &= \mathcal{F}^{-1} \left[ \left( \Tilde{U}_{\text{sp}} \left( \bm{u} \right) \ast \Tilde{o} \left( \bm{u} \right) \right) \cdot P \left( \bm{u} \right) \right],\label{eq:speckleSIM}
\end{split}
\end{equation}
where $\ast$ denotes convolution operation. The speckle SIM can encode a frequency bandwidth of $\left( \text{NA}_{\text{obj}} + \text{NA}_\text{speckle} \right)/ \lambda$, where $\text{NA}_\text{speckle}$ is the effective NA of the speckle illumination. Similar to sinusoidal SIM, each raw measurement of speckle SIM is also band-limited by $P$, and multiple raw measurements are needed to decode the super-resolved information. In practice, a varying speckle illumination is often needed to collect raw measurements for a well-conditioned super-resolution reconstruction~\cite{mudry2012structured, yeh2017structured}.

% static speckle + object dynamic
In Speckle Flow SIM, we use a single static speckle illumination pattern and instead rely on the inherent motion dynamics of the scene to diversify the measured information (Fig.~\ref{fig:SIMintro}(d)). We acquire a sequence of frames for a dynamic scene that can be represented as $o\left( \bm{r}, t \right) = \textit{motion}\left( o\left( \bm{r} \right), \bm{\mu}\left(\bm{r}, t \right) \right)$. $\textit{motion}\left( \cdot, \cdot \right)$ is a motion function transforming a time-independent scene, $o\left( \bm{r} \right)$, to a timepoint by its spatially-varying motion kernel, $\bm{\mu}\left(\bm{r}, t \right)$. Speckle Flow SIM can be formally expressed as
\begin{equation}
\begin{split}
    U&_{\text{SpeckleFlow}} \left( \bm{r}, t \right) \\
    &= \mathcal{F}^{-1} \left[ \mathcal{F} \left( U_{\text{sp}}\left( \bm{r} \right) \cdot  \textit{motion}\left( o\left( \bm{r} \right), \bm{\mu}\left(\bm{r}, t \right) \right) \right) \cdot P\left( \bm{u} \right) \right].\label{eq:SpeckleFlow}    
\end{split}
\end{equation}

\begin{figure}[]
\centering
\includegraphics[width=0.6\linewidth]{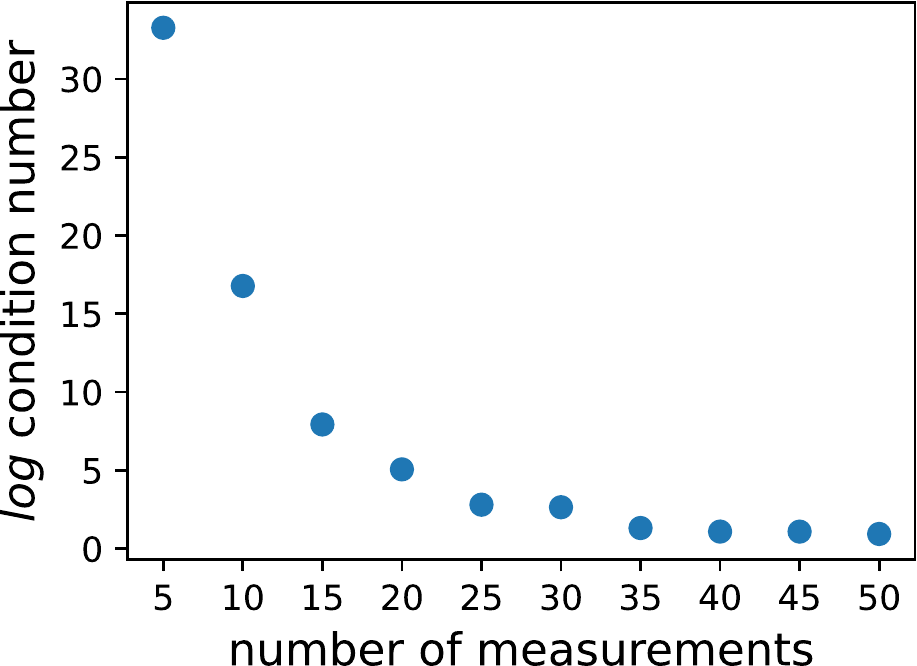}
\caption{Condition number analysis for Speckle Flow SIM in 1D with increasing numbers of raw measurements. This suggests that Speckle Flow SIM becomes well-posed with a sufficient number of raw measurements.}
\label{fig:condition_num}
\end{figure}

It is difficult to analyze the well-posedness of the Speckle Flow SIM inverse problem for deformable motion. Hence, we show here only a simplified analysis for the case of translational motion, and assume the result will be similar for deformable motion. In the translational motion case, $\textit{motion}\left( o\left( \bm{r} \right), \bm{\mu}\left(\bm{r}, t \right) \right) = o\left( \bm{r} + \delta \bm{r}_t \right)$, where $\delta \bm{r}_t$ is the relative displacement at timepoint $t$, and thus Eq.~\ref{eq:SpeckleFlow} can be written as a linear system:
\begin{equation}
\begin{split}
    U&_{\text{SpeckleFlow}} \left( \bm{r}, t \right) = \mathcal{F}^{-1} \left[ \mathcal{F} \left( U_{\text{sp}}\left( \bm{r} \right) \cdot   o\left( \bm{r} + \delta \bm{r}_t \right) \right) \cdot P\left( \bm{u} \right) \right] \\
    &= \mathcal{F}^{-1} \left[ \left( \Tilde{U}_{\text{sp}} \left( \bm{u} \right) \ast \left( \Tilde{o} \left( \bm{u} \right) \cdot e^{-2\pi i \delta \bm{r}_t \cdot \bm{u}} \right) \right) \cdot P \left( \bm{u} \right) \right].\label{eq:SpeckleFlow_linear}
\end{split}
\end{equation}
Because of the translation property of Fourier transforms, the motion kernel in Fourier space can be expressed as $e^{-2\pi i \delta \bm{r}_t \cdot \bm{u}}$. If we rewrite this linear forward model into a transformation matrix, the transformation matrix changes at different timepoints as the scene moves. 
We then perform a condition number analysis for using this forward model to solve for 2$\times$ super-resolution. In order to make the singular value decomposition (SVD) of the condition number calculation computationally feasible, we only consider the 1D case, where the scene and the measurement are both one-dimensional. The results (Fig.~\ref{fig:condition_num}) show that the condition number of Speckle Flow SIM is dependent on the number of raw measurements captured, and the problem is better posed as more raw measurements are added, as expected. Beyond having a sufficient number of measurements, the super-resolution reconstruction is well-posed and the condition number becomes asymptotic.

\subsection{Background: Coordinate-based Neural Networks}
\label{sec:coord_based}

% MLP, input-output, algorithm, implication
% maybe should use r consistently (instead of xy)
The coordinate-based neural network is an alternative representation of a grid-based matrix using a MLP. The coordinate-based neural network takes an arbitrary coordinate of the matrix as the input and outputs the corresponding matrix value. To represent a 2D scene, a coordinate $\bm{r}=\left(x, y\right)$ within the domain of interest is used as the input coordinate. The weight, $\theta$, of the MLP, $f$, is optimized to fit into the given matrix~\cite{sitzmann2020implicit, martel2021acorn}, such that
\begin{equation}
    \argmin_{\theta} \sum_{\bm{r} \in \textit{domain}\left( o \right)} \left| f\left(\bm{r}; \theta \right) - o\left(\bm{r}\right) \right|^2. \label{eq:MLP_background}
\end{equation}
Compared with a matrix representation, the coordinate-based MLP has a continuous form without the matrix grid, such that any off-grid coordinate values can be queried without additional rounding or interpolation. Once the MLP weights are optimized, we can retrieve a matrix with an arbitrary sampling grid by querying all the corresponding coordinates from the MLP. The coordinate-based MLP also tends to have a smoothing effect on the retrieved matrix because of the linearity of its fully-connected layers~\cite{tancik2020fourier}.

% positional encoding
Positional encoding maps a coordinate into a vector of sinusoidal features at different frequencies before feeding into the network. This helps avoid over-smoothing and enables the representation of high-frequency details~\cite{tancik2020fourier}. Positional encoding, $\gamma$, can be formally written as
\begin{equation}
    \gamma \left( \bm{r} \right) = \left( \bm{r}, \cos{\left(2^i \pi \bm{r} \right)}, \sin{\left(2^i \pi \bm{r} \right)}, ... \right), \text{for } i=0, ..., l-1. \label{eq:posenc}
\end{equation}
$l$ is the order of positional encoding, which is a tunable parameter. As a result, positional encoding maps $\bm{r} \in \mathbb{R}^2$ to $\gamma \left( \bm{r} \right) \in \mathbb{R}^{4l+2}$ as the input for the coordinate-based MLP.

\subsection{Neural Space-Time Model}
\label{sec:spacetime_model}

% intuition/assumption behind
The neural space-time model is a compressive representation of a dynamic scene. The dynamic scene is split into two parts under the neural space-time model: motion kernels corresponding to different timepoints stored in the motion MLP, $f_{\textit{motion}}$, and a time-independent scene represented by the scene MLP, $f_{\textit{scene}}$. Both MLPs are coordinate-based.

The motion MLP acts as the motion kernel in Eq.~\ref{eq:SpeckleFlow}, which transforms a dynamic scene into a time-independent scene. For any space-time coordinate, $\left( \bm{r}, t \right)$, the motion MLP estimates the relative displacement with respect to the scene captured by the scene MLP, $\delta \bm{r}_t$. The dynamic scene can be expressed using the motion MLP, namely,
\begin{equation}
o\left( \bm{r}, t \right) =  o \left( \bm{r} + \delta \bm{r}_t \right) = o \left( \bm{r} + f_{\textit{motion}} \left( \bm{r}, t ; \theta_{\textit{motion}} \right) \right),
\end{equation}
where $\theta_{\textit{motion}}$ is the weights of the motion MLP. Since the relative displacement returned from the motion MLP can vary spatially, the motion MLP can represent both global motion and locally deformable motion dynamics. When the motion MLP is queried for every spatial coordinate at a given timepoint, the obtained motion kernel can be used to map the current scene to the time-independent scene.

The scene MLP captures a time-independent, super-resolved scene. I.e., we can feed in any spatial coordinate to the scene MLP and obtain its corresponding value from the MLP output, $\hat{o}\left( \bm{r} \right) =  f_{\textit{scene}} \left( \bm{r}; \theta_{\textit{scene}} \right) $, where $\theta_{\textit{scene}}$ is the weights of the scene MLP. The scene MLP does not take the timepoint as an input, as the time-dependency is already account in the motion MLP. We use $\hat{o}$ here since the scene MLP output is an approximated value from the coordinate-based MLP network, which might not be exact. 

% chain together two mlps
Combining these two parts together, as shown in Fig.~\ref{fig:overview}(b), we can obtain the pixel value at any spatial and temporal coordinate by querying the estimated motion from the motion MLP first and then using the motion-accounted spatial coordinates to retrieve the super-resolved scene from the scene MLP. The positional encoding is applied to the motion-accounted coordinate before feeding into the time-independent scene MLP. The final approximated scene can be expressed as 
\begin{equation}
    \hat{o}\left( \bm{r}, t; \theta_{\textit{motion}}, \theta_{\textit{scene}} \right) = f_{\textit{scene}} \left( \gamma \left( \bm{r} + f_{\textit{motion}} \left( \bm{r}, t ; \theta_{\textit{motion}} \right) \right); \theta_{\textit{scene}} \right).
\end{equation}
We repeat this process for each pixel of our sampling frame to retrieve a scene at a given timepoint, $\hat{o}\left(t; \theta_{\textit{motion}}, \theta_{\textit{scene}} \right)$. We can input the retrieved scene into the forward model as Eq.~\ref{eq:SpeckleFlow} to simulate the raw image captured by the camera.

\subsection{Dynamic Scene Reconstruction}
\label{sec:recon}
% optimization math expression here
The neural space-time model recovers a dynamic scene from the weights of the motion and the scene MLPs. During the reconstruction, the model weights are optimized to minimize the loss function, which is the difference between the acquired intensity image at timepoint $t$, $I_t$, and the simulated intensity image using the forward model described in Sec.~\ref{sec:spacetime_model}. This optimization can be formulated as
\begin{equation}
    \argmin_{\theta_{\textit{motion}}, \theta_{\textit{scene}}} \sum_t \left\| \sqrt{I_t} - \left| \mathcal{F}^{-1} \left[ \mathcal{F} \left( U_{\text{sp}} \cdot  \hat{o}\left(t; \theta_{\textit{motion}}, \theta_{\textit{scene}} \right) \right)  \cdot P \right] \right| \right\|^2. \label{eq:reconstruction}
\end{equation}
We drop the spatial coordinates, $\bm{r}$ and $\bm{u}$, in this expression for simplicity. The complex-field of the speckle illumination, $ U_{\text{sp}}$, is predetermined in a separate calibration process (described in Sec.~\ref{sec:speckle_calibration}) before the reconstruction. As both the forward model and the MLPs are differentiable, the gradients for those two MLPs' weights can be computed through back-propagation, and we then update the weights, $\theta_{\textit{motion}}$ and $\theta_{\textit{scene}}$, using gradient descent. We also note that $I_t$ can be replaced by the measured complex-field using an interferometric-based system, such as in~\cite{chowdhury2013structured}.

% additional term on loss function (high-freq suppression)
While the reconstruction does not require any regularization terms, common regularizers for matrix-based inverse problem solution (e.g., L1, Tikhonov, total variation, etc.) are also available for our reconstruction. However, unlike in the matrix representation, applying a spatial filter directly to update the weights of a coordinate-based MLP is difficult. The experimental data reconstruction often generates some very high-frequency signals caused by imaging noise. To filter out these from the model weights, we have a high-frequency suppression term, $\mathcal{L}_{\textit{high-freq}}$, to penalize the reconstruction of the signals beyond the theoretical super-resolution limit of Speckle Flow SIM. I.e., $\mathcal{L}_{\textit{high-freq}} = \left\| \mathcal{F} \left( o\left( t \right) \right) \left( 1 - P^\prime \right) \right\|_1$, where $P^\prime$ is the pupil function for the effective NA of Speckle SIM, $\text{NA}_{\textit{speckle}} + \text{NA}_{\textit{obj}}$.

\begin{figure*}[]
\centering
\includegraphics[width=\linewidth]{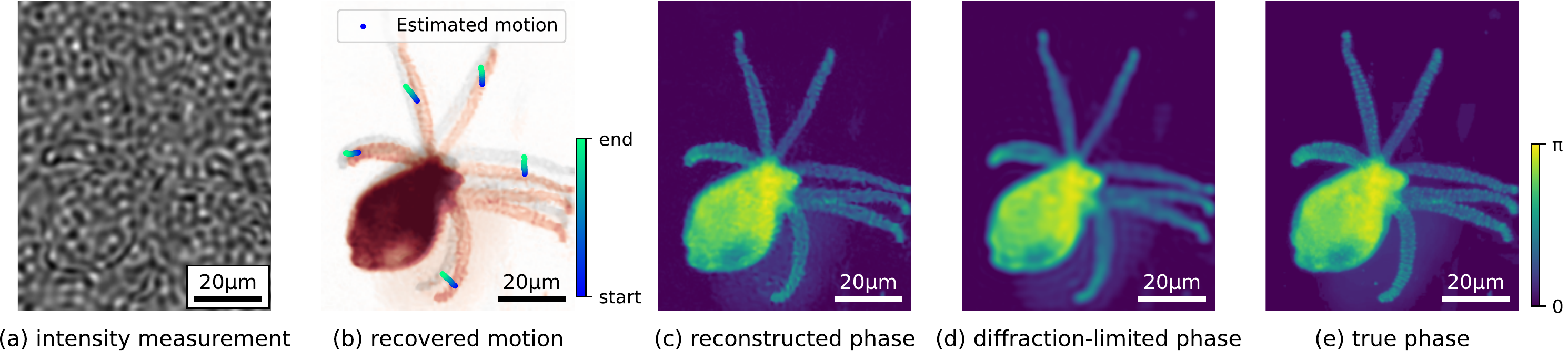}
\caption{Simulation results for Speckle Flow SIM dynamic scene reconstruction of hydra with deformable motion. (a) The first frame of intensity measurements. (b) Recovered deformable motion trajectories for selected points are drawn as color gradient lines, where a point's color indicates its corresponding timepoint and the first (red) and last frame (grey) of the reconstruction are overlaid. (c) The reconstructed phase at the first timepoint of the dynamic scene. (d) Diffraction-limited phase obtained by low-pass filtering from (e) the true phase. The dynamic reconstruction is in Supplementary Video 1.}
\label{fig:simu_hydra}
\end{figure*}

\section{Implementation}
\label{sec:implementation}

\subsection{Neural Space-Time Model Setup}
\label{sec:model_details}
% network architecture details
The dynamic scene reconstruction is computation and memory intensive. As the scene is stored in the coordinate-based neural networks, obtaining the value for each pixel of the scene takes a query to the neural space-time model with its coordinate. Since the forward model requires the full matrix of a scene to perform a Fourier transform, we need to repeat this query for an entire scene (usually at the scale of a million pixels) to simulate a measured image. Besides, the model's intermediate output values for each query are stored in the memory for an efficient gradient computation, and thus numerous copies of the intermediate output need to fit into the memory for the reconstruction in Eq.~\ref{eq:reconstruction}. To make this computationally feasible, we use a compact network configuration for the motion and scene MLPs. The motion MLP has the network depth of 4 and width of 32. While the spatial coordinates are fed into the motion MLP directly, the time coordinate is encoded with a positional encoding order of 4 for the motion MLP's input as in~\cite{park2021nerfies}. The scene MLP has a network depth of 8 and width of 64, which is larger than the motion MLP as we expect the scene to contain more information than the motion. The scene MLP uses a skip connect to concatenate the input to the fifth layer's output as in~\cite{mildenhall2020nerf}. As the MLP outputs real values, the scene MLP has two output channels for phase and absorption of the scene respectively. Both MLPs use ReLU activation function after each fully-connected layer. The reconstruction using our network configuration can be performed in a single Titan Xp GPU (Nvidia) with 12 GB memory. Without an exhaustive testing of other configurations due to our limited computational resources, we also find our configuration robust for different scenes or dynamics. Nevertheless, having larger MLPs might still help to represent a more complex scene and motion as suggested in the universal function approximator theory~\cite{hornik1989multilayer}.

\subsection{Reconstruction Procedure}
\label{sec:recon_proc}
We implement the neural space-time model and the reconstruction using Jax library (Google)~\cite{bradbury2018jax} and in-house developed computational imaging toolbox\footnote{Code will be released upon publication.}. We use ADAM, a fast version of the stochastic gradient descent, for the optimization of the neural space-time model. The motion and scene MLPs are updated concurrently under the same setting. The learning rate is set to $5\times 10^{-4}$ for simulation and $5\times 10^{-5}$ for experimental data, with a exponential delay to $0.1$ of the starting value at the end. The reconstruction process takes 200k update steps for simulation data and 500k steps for experimental data. We only optimize for the data fidelity term as in Eq.~\ref{eq:reconstruction} without the high-frequency suppression or other regularization terms for simulation data. For experimental data, we tried three different settings: vanilla reconstruction (no regularization), the reconstruction with high-frequency suppression, and the reconstruction with high-frequency suppression and speckle update (described in Sec.~\ref{sec:speckle_calibration}). The high-frequency suppression has a weighting factor of $1\times 10^{-5}$ when included.

\subsection{Simulation Setup}
\label{sec:simu_setup}
We first perform a simulation study to validate Speckle Flow SIM. A speckle illumination is generated by a plane wave passing through a thin, random phase mask, and we low-pass filter the phase mask such that $\text{NA}_{\textit{speckle}} = \text{NA}_{\textit{obj}}$. The field of the speckle illumination is known during the reconstruction of the simulation data. We use two phase phantoms in the simulation: the Shepp-Logan phantom with a rigid motion we define using translation and rotation (40 frames total), and a video of hydra with a deformable motion (20 frames total). Then, we simulate a sequence of measurements frame-by-frame using the forward model of speckle SIM as in Eq.~\ref{eq:speckleSIM}. The simulated sequence of intensity images is used as the input for the Speckle Flow SIM reconstruction.

\subsection{Experiment Setup}
% imaging system
We build a custom microscope system as Fig.~\ref{fig:overview}(a). A 532 nm green laser light (Thorlabs CPS532, 4.5 mW) outcoming from a single-mode fiber is collimated into a plane wave which then shines on a random diffuser for the speckle illumination. We use four layers of Scotch tape (3M 810 Scotch Tape, S-9783) as the random diffuser~\cite{yeh2019computational}. The layered Scotch tape is attached on a clear microscope slide for enhanced stability. After propagating through the sample, the transmitted light is magnified by a 4$f$ system composed of a $10\times$ 0.25NA objective lens (Nikon) and a single lens (250 mm focal length) as the tube lens, resulting in an effective magnification of 12.5. The magnified image is captured by a monochromatic CMOS sensor (FLIR, BFS-U3-200S6M-C) placed at the back focal plane of the tube lens. The exposure time is set to maximize the dynamic range of the sensor.

% positional encoding
\begin{figure*}[]
\centering
\includegraphics[width=\linewidth]{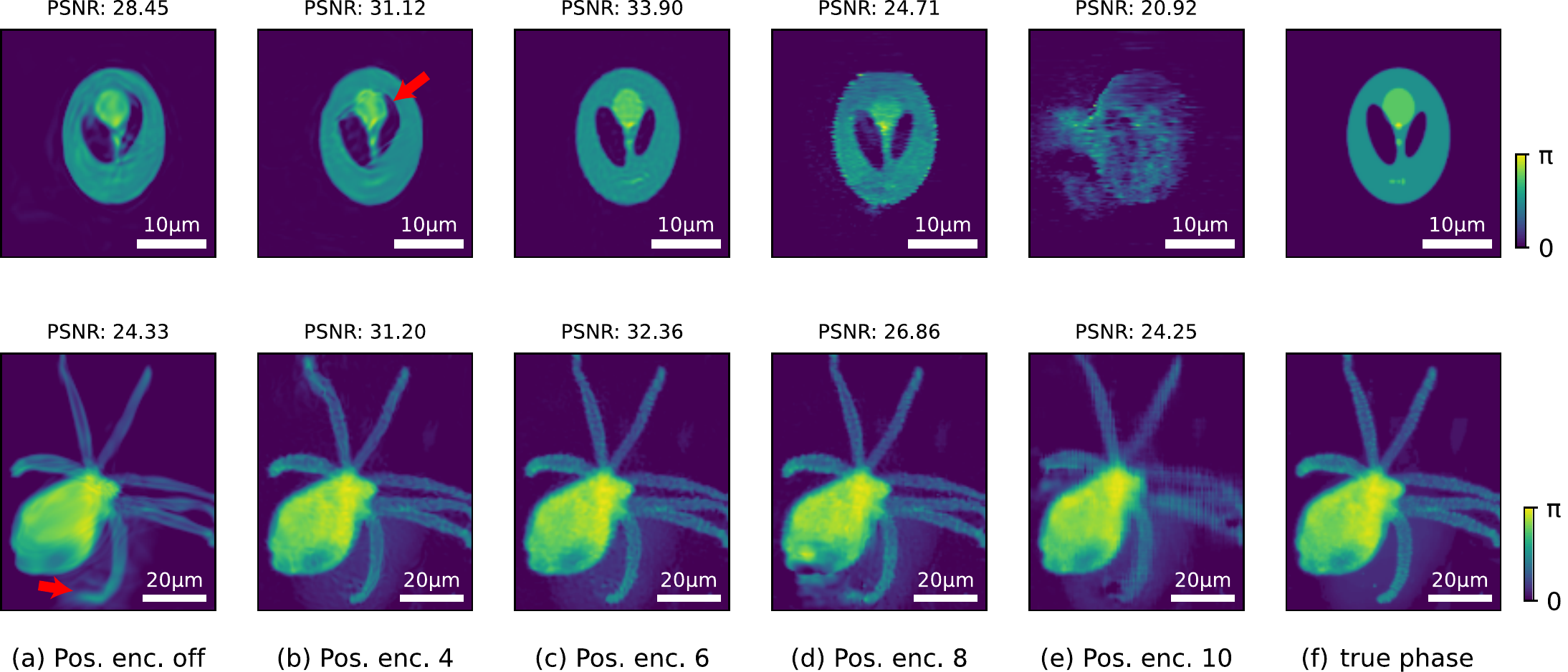}
\caption{Dynamic scene reconstructions for Shepp-Logan phantom and hydra phantom with different positional encoding orders. The phase reconstruction at the first timepoint of the sequence is shown. The peak signal-to-noise ratio (PSNR) is calculated for each reconstructed sequence. The red arrows indicate the distortions caused by the inexact motion estimation. The dynamic reconstruction is in Supplementary Video 2.}
\label{fig:simu_posenc}
\end{figure*}

\subsection{Speckle Calibration}
\label{sec:speckle_calibration}
We need to calibrate for the complex field of the speckle illumination in advance before the experimental data reconstruction. The calibration can be performed by simply imaging the background in a holographic setup. In our intensity-only system, however, we retrieve the phase of the speckle illumination by taking intensity images at different defocus planes~\cite{allen2001phase}. The intensity image at each defocus plane, $I_{z_i}$, is modeled as the in-focus speckle field propagating by a distance of $z_i$ using the angular spectrum method~\cite{goodman2005introduction}, such that
% modeling
\begin{equation}
    I_{z_i}\left( \bm{r}\right) = \left| \mathcal{F}^{-1} \left( \Tilde{U}_{\textit{sp}} \left( \bm{u}\right) \cdot  P\left( \bm{u}\right) \cdot \exp{\left(j \frac{2\pi z_i \sqrt{1 - \bm{u}^2}}{\lambda} \right)} \right) \right|^2. \label{eq:calibration}
\end{equation}
We then solve for $U_{\textit{sp}}$ from the acquired defocused images $I_{z_i}$.

In the experimental setup, we place the CMOS sensor on a 1-axis motorized stage (Thorlabs Z825) toward the $z$ direction. To retrieve the phase of the speckle illumination, we acquire intensity images at 10 $z$-planes in the image space with a step size of 200 µm, which is equivalent to 1.28 µm in the sample space. This defocus calibration process is performed on the background field-of-view without the scene. After the acquisition, the complex field of the speckle illumination is iteratively updated to minimize Eq.~\ref{eq:calibration} using gradient descent. 

The speckle field retrieved from this calibration process may not perfectly match the actual speckle illumination in the dynamic scene acquisition for experimental reasons, such as mechanical instability. Thus, the complex field of the speckle can also be jointly updated to minimize the objective function defined in Eq.~\ref{eq:reconstruction} during the reconstruction of a dynamic scene. As the speckle update affects the scene reconstruction, the speckle update is performed only in the fine-tuning stage of the dynamic scene reconstruction, which is after the first 10k update steps in our case.

\section{Results}
\label{sec:results}

\subsection{Simulation Results}
\label{sec:simulation_results}

% deformable motion (hydra)
We first validate the reconstruction of the hydra phantom with deformable motion in simulation. As the hydra phantom in our simulation is phase-only, the simulated intensity measurement in Fig.~\ref{fig:simu_hydra}(a) looks similar to the speckle image without the sample and does not contain much discernible contrast without the reconstruction. After the reconstruction, the motion trajectories recovered by the neural space-time model are a good fit for the actual scene dynamics as in Fig.~\ref{fig:simu_hydra}(b) and Supplementary Video 1, which demonstrates the motion MLP's capacity of representing deformable motion. The phase reconstruction for the first frame of the scene is shown in Fig.~\ref{fig:simu_hydra}(c). Compared with the diffraction-limited reference scene in Fig.~\ref{fig:simu_hydra}(d) (low-pass filtered from the true scene), the reconstruction recovers those finer features on both the hydra's gastrovascular cavity and tentacles, matching well to the original ground truth phantom. It is worth noting that the space-time model here successfully reconstructs the discontinuous and non-smooth features from the scene MLP, e.g. fine features on the tentacles, which can be credited to the non-linearity from the activation function and the high-frequency mapping from positional encoding.

% number of frames
\begin{figure}[]
\centering
\includegraphics[width=\linewidth]{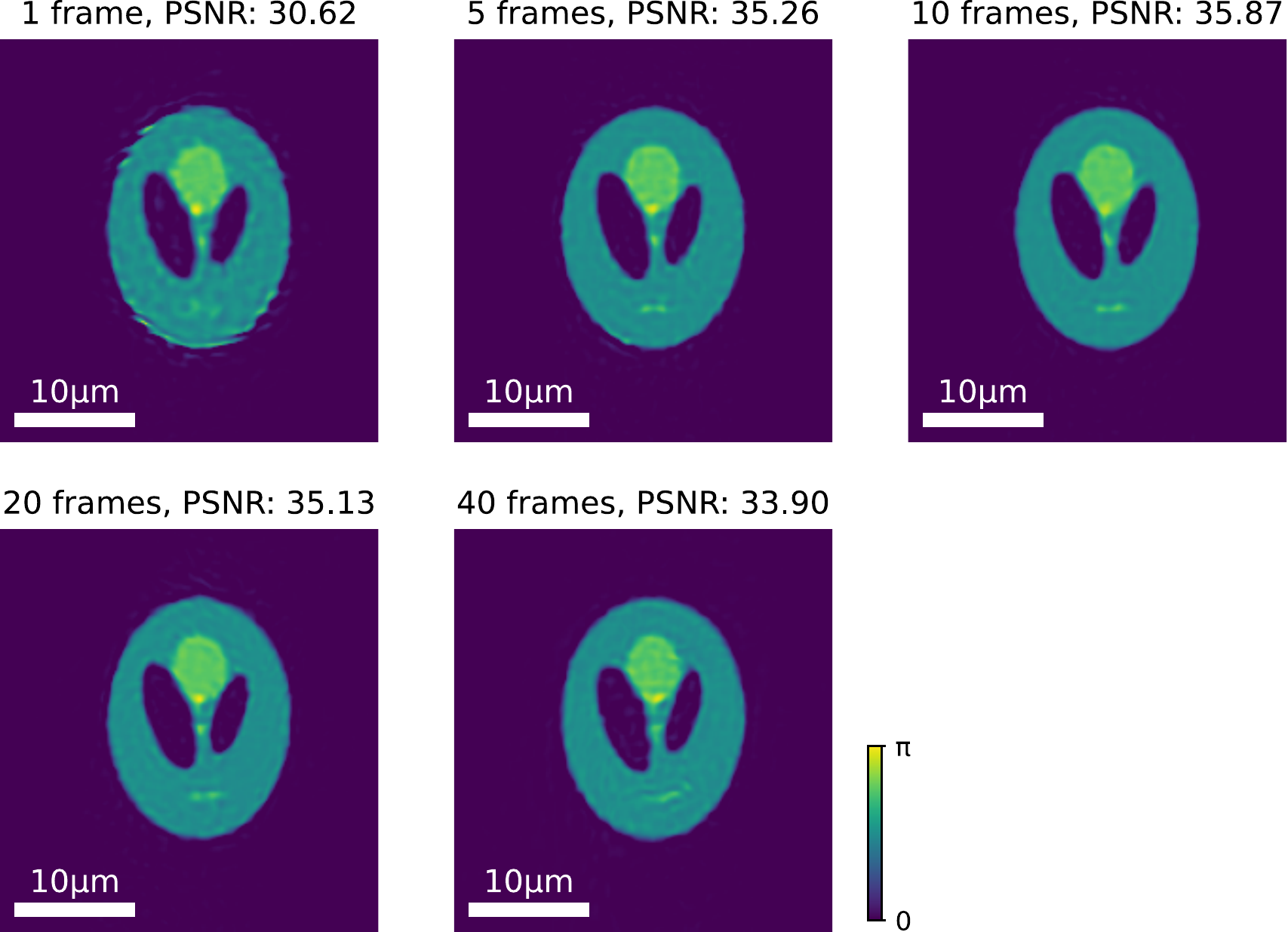}
\caption{The phase reconstructed using the first 1, 5, 10, 20, 40 frames of the input intensity image sequence. The reconstructed phase at the first timepoint is shown here. The peak signal-to-noise ratio (PSNR) is calculated for the reconstruction over all timepoints. Based on the PSNR, the reconstruction quality is optimal using 10 frames.}
\label{fig:simu_num_frames}
\end{figure}

We further analyze the effect of positional encoding on the reconstruction performance. By its definition in Eq.~\ref{eq:posenc}, the order of positional encoding, $l$, regulates the highest frequency of the positional encoding mapping, which in turn affects the reconstruction of high-frequency signal~\cite{tancik2020fourier}. With the same set of input raw measurements and reconstruction settings, we reconstruct the dynamic scene with different orders of positional encoding of the spatial coordinates in the scene MLP, as shown in Fig.~\ref{fig:simu_posenc}. When the positional encoding is turned off or the order is low (order of 4), the reconstructions are under-fitted and also have distortions as indicated by red arrows. The distortions are caused by the inexact motion estimation for some small features (Supplementary Video 2). The positional encoding order of 6 gives the optimal reconstruction in terms of its peak signal-to-noise ratio (PSNR) and visual quality for both phantoms. When the order is set to 10, the reconstructed phase contains a considerable scene distortion and high-frequency artifacts. This suggests that while a high order of positional encoding gives the scene MLP extra degrees-of-freedom for the scene representation, it hinders the convergence of the motion MLP at the same time. The scene MLP with a high order of positional encoding tends to over-fit high-frequency content, before a good motion estimation is obtained by the motion MLP.

We also compare the reconstructions with different numbers of input intensity images, such that we reconstruct using the first 1, 5, 10, 20, 40 images of the simulated measurement sequence for the Shepp-Logan phantom. With more input images, the neural space-time model receives more encoded information of the scene, while it is also responsible for recovering the motion dynamics at more timepoints. As in Fig.~\ref{fig:simu_num_frames}, despite being ill-posed, the phase can be reconstructed from a single frame due to the implicit network smoothness prior from the scene MLP. The reconstruction improves with more raw images and reaches the optimal quality using 10 frames. If even more frames are used, the reconstruction slowly degrades as the motion MLP has to recover an extended scene dynamic from an increasing number of timepoints. The optimality of 10 raw images roughly coincide with the number of raw images required for sinusoidal SIM.

\subsection{Experimental Results}
\label{sec:experiment_results}

\begin{figure*}[]
\centering
\includegraphics[width=0.9\textwidth]{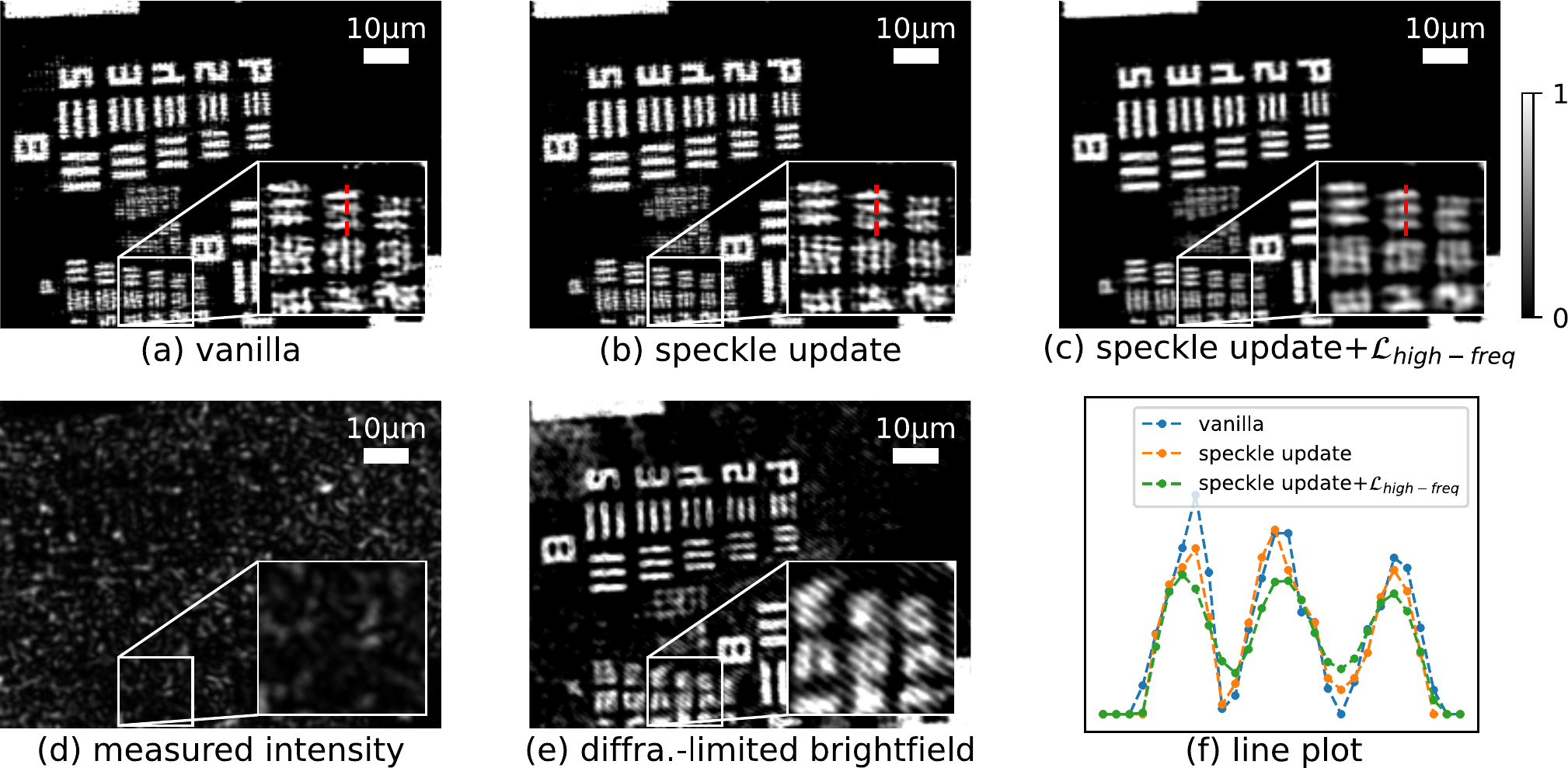}
\caption{The experimental reconstruction of an amplitude USAF-1951 resolution target in continuous motion. (a)-(c) The reconstructed absorption coefficient at the first timepoint under different reconstruction settings described in Sec.~\ref{sec:recon_proc}. (d) The corresponding raw intensity image. (e) The diffraction-limited brightfield image as a reference. (f) Line plot for the red dotted lines in (a)-(c). The dynamic reconstruction is in Supplementary Video 3.}
\label{fig:experiment_result}
\end{figure*}

As a proof-of-concept experiment for Speckle Flow SIM, we create a dynamic scene by placing an amplitude USAF-1951 resolution target (Benchmark Technologies) on a 1-axis motorized stage (Thorlabs Z812) that travels laterally ($xy$-direction). A sequence of 40 intensity images is acquired while the resolution target moves continuously. The defocus images for speckle calibration are captured before the actual image acquisition. As the speckle evolves with a different optical path length, we calibrate the speckle using a background field-of-view on the resolution target slide to minimize the potential mismatch. A diffraction-limited brightfield image is also acquired using the same system without the scattering layer as a reference.

% three recon settings
The reconstruction is performed with three different settings detailed in Sec.~\ref{sec:recon_proc}. The reconstruction is shown in Fig.~\ref{fig:experiment_result}(a)-(c). The brightfield image is shown in Fig.~\ref{fig:experiment_result}(e) for comparison. The vanilla reconstruction recovers the high-frequency information well despite being noisy, as demonstrated on the zoom-in and the line plot in Fig.~\ref{fig:experiment_result}(f). The noisy reconstruction is due to the mismatch between the calibrated speckle and the actual speckle for practical reasons, such as, subtle differences in optical path length, mechanical vibration and instability. Adding a joint speckle update reduces the noise in the reconstruction while still maintaining good contrast for fine features (Fig.~\ref{fig:experiment_result}(b)). When we add the high-frequency suppression regularization term, described in Sec.~\ref{sec:recon}, to the reconstruction loss, the reconstruction in Fig.~\ref{fig:experiment_result}(c) becomes much smoother and less noisy at the cost of slightly reduced contrast.

% resolution gain assessment
The diffraction-limited image has a theoretical Rayleigh resolution of $1.22 \cdot \lambda /\text{NA} = 2.60\text{ µm}$, which corresponds to Group 8 Element 5 of the USAF resolution target (2.46 µm) and matches our observation in Fig.~\ref{fig:experiment_result}(e). In Speckle Flow SIM, our reconstruction resolves up to group 9 element 4 (1.38 µm), which is $1.88\times$ of the diffraction-limited resolution. Since the speckle calibration is performed with the same objective lens such that $\text{NA}_{speckle} \leq \text{NA}_{obj}$, we expect a theoretical resolution gain of $2\times$, very close to our experimental demonstration. This experimental resolution improvement is similar to the reported gain from previous study of sinusoidal SIM~\cite{gustafsson2000surpassing} and speckle SIM~\cite{mudry2012structured} which are both close to 2$\times$.

\begin{figure*}[]
\centering
\includegraphics[width=\textwidth]{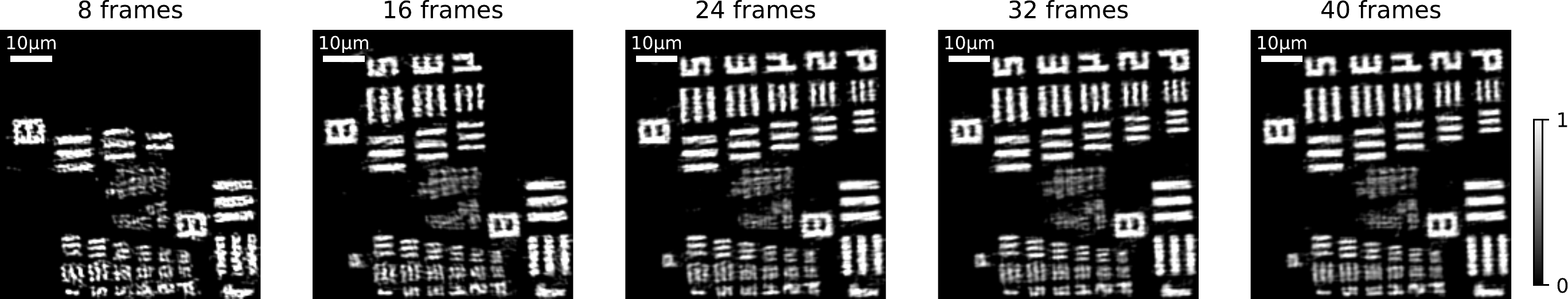}
\caption{The experimental reconstruction of the moving USAF-1951 target using the first 8, 16, 24, 32, 40 frames of the acquired intensity image sequence. The reconstructed absorption coefficient at the first timepoint is shown here.}
\label{fig:experiment_num_frames}
\end{figure*}
Similar to Sec.~\ref{sec:simulation_results}, we compare the reconstructions using different numbers of acquired intensity images. Fig.~\ref{fig:experiment_num_frames} shows the reconstruction using the first 8, 16, 24, 32, 40 acquired frames. The reconstruction is visually identical using 24, 32, 40 acquired frames, while some regions of the scene are missed when reconstructing with 8 or 16 frames. This is caused by the joint speckle update, such that the static speckle background becomes indistinguishable from the dynamic scene foreground when the number of raw frames is limited.

\subsection{Beyond $2\times$ Super-resolution}

Speckle SIM can achieve more than $2\times$ better than diffraction-limited resolution when the illumination NA is higher than the objective NA, such that the final resolution is $\frac{\lambda}{\text{NA}_{\text{obj}} + \text{NA}_{\text{speckle}} } > 2 \cdot \frac{\lambda}{\text{NA}_{\text{obj}}}$, as in Sec.~\ref{sec:theory}. This setting can be useful for high-content imaging~\cite{yeh2019computational, yeh2019speckle} or for total internal reflection-based SIM settings~\cite{cragg2000lateral}. Speckle Flow SIM can also similarly recover a dynamic scene with beyond $2\times$ diffraction-limited resolution using a low-NA system and a fine speckle illumination.

\begin{figure*}[]
\centering
\includegraphics[width=\textwidth]{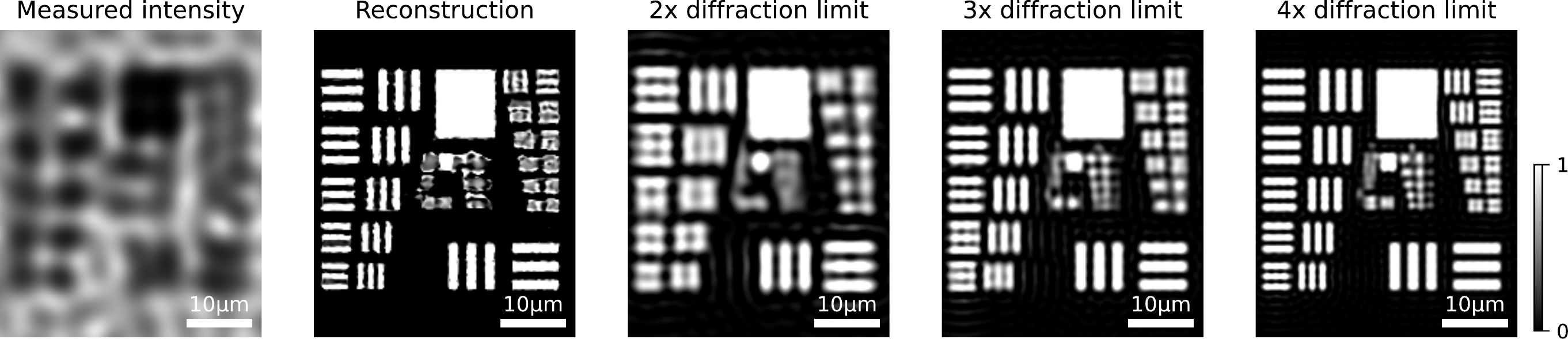}
\caption{The reconstructed absorption coefficient for an amplitude USAF-1951 resolution target in motion achieves more than $2\times$ better resolution than the diffraction limit in simulation. The numerical aperture (NA) of the speckle is 3$\times$ the NA of the objective lens. The target with $2\times$, $3\times$, and $4\times$ the diffraction-limited resolution are shown as references. The reconstruction is close to $4\times$ the diffraction-limited resolution.}
\label{fig:simu_more_res}
\end{figure*}

We validate this in simulation, for an objective lens of 0.1 NA and known speckle illumination of 0.3 NA. The amplitude USAF-1951 resolution target with a constant-velocity translational motion is imaged, and we acquire the intensity images at 100 equally-spaced timepoints. The reconstruction is performed under the same optimization procedure as in Sec.~\ref{sec:recon_proc}. The reconstructed absorption coefficient as well as a few reference targets are shown in Fig.~\ref{fig:simu_more_res}. The reference targets are obtained by directly low-pass filtering from the groundtruth target with the bandwidth of $2\times$, $3\times$, and $4\times$ the diffraction limit. The reconstruction is able to resolve the second from the top element on the right group, which is close to the $4\times$ diffraction-limited reference in Fig.~\ref{fig:simu_more_res}, matching well with the theoretical limit in Sec.~\ref{sec:theory}. 

\section{Limitations}
\label{sec:limitations}

% computation load
Speckle Flow SIM has several limitations. First, the reconstruction process is computationally expensive and memory intensive as discussed in Sec.~\ref{sec:model_details}. The 500k iterations of experimental data reconstruction in Fig.~\ref{fig:experiment_result} (matrix size: $400\times 540$) takes around 13 hours on a Nvidia Titan Xp GPU using our current implementation. This reconstruction time is very long compared with other SIM methods for static scenes. E.g., the reconstruction of sinusoidal SIM takes 30s using one CPU core~\cite{gustafsson2000surpassing}); speckle SIM takes 5 hours using one CPU core~\cite{mudry2012structured} and 15 minutes on a GPU~\cite{yeh2019computational}). The main bottleneck for the reconstruction time is the coordinate-based MLPs in the neural space-time model, which requires one MLP's forward pass to know the corresponding value for each pixel coordinate. Speckle Flow SIM may benefit from two very recent studies suggesting a potential 100x speed-up with a more efficient sparse voxel representation and GPU programming~\cite{yu2021plenoxels, mueller2022instant}.

\begin{figure*}[]
\centering
\includegraphics[width=0.75\textwidth]{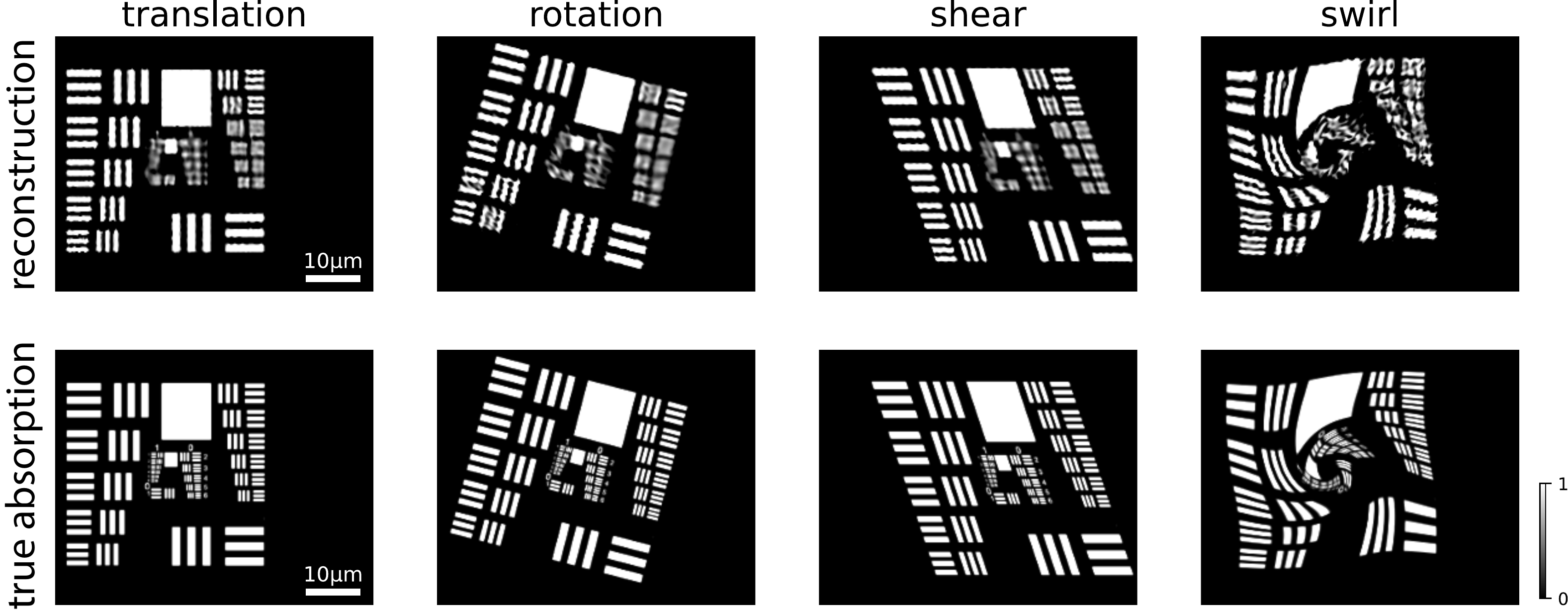}
\caption{The reconstructed absorption coefficient for an amplitude USAF-1951 resolution target with four different types of motion in simulation. The performance of Speckle Flow SIM is motion-dependent and degrades with highly-deformable motion.}
\label{fig:simu_motion_comparison}
\end{figure*}

% motion assumption, smoothness prior
Second, the neural space-time model assumes that the scene at all timepoints can be wrapped into the time-independent frame represented by the scene MLP. This assumption is necessary to exploit the smoothness of motion using the motion MLP for a joint motion-scene optimization and improve the SIM frame rate by an order of magnitude. However, this assumption also limits the applicability of Speckle Flow SIM, such that dynamics must be related by motion estimations, e.g., cells moving a microfluidic chamber. Speckle Flow SIM will fail to capture those non-smooth dynamics, e.g., random firing of neurons. Besides, the overall performance of Speckle Flow SIM is also motion-dependent. In Fig.~\ref{fig:simu_motion_comparison}, we reconstruct the USAF-1951 resolution target with four different types of motion dynamics (i.e., translation, rotation, shearing, and swirl) in simulation. The forward simulation and reconstruction settings are the same as in Sec.~\ref{sec:simu_setup}. As in Fig.~\ref{fig:simu_motion_comparison}, the reconstruction of the swirl motion contains more noise or artifacts than those of affine motion. This suggests that the performance of our method degrades for complex motion dynamics (e.g., highly deformable motion), presumably due to the inexact motion estimation from the motion MLP. As also discussed in Sec.~\ref{sec:simulation_results} and Fig.~\ref{fig:simu_num_frames}, the representation capacity of a motion MLP is finite and limited, and thus a motion MLP with a compact network architecture may fail to fit into dramatic or highly deformable motion dynamics. A larger network architecture may help accommodate the estimation for more complex motion dynamics~\cite{hornik1989multilayer}.

% calibration process
Third, Speckle Flow SIM requires a calibration process of the speckle illumination before the reconstruction, which is unlike previous speckle SIM methods. Previous methods either assumed a statistical prior on the speckle~\cite{mudry2012structured, yeh2017structured, mangeat2021super} or jointly resolved the speckle without any prior assumption~\cite{yeh2019computational, yeh2019speckle}. Having the full knowledge of the random speckle illumination allows Speckle Flow SIM to super-resolve a scene using much fewer raw images than other methods. Our current calibration process, however, limits Speckle Flow SIM to $2\times$ the diffraction-limited resolution, as the calibration images are also captured in the same optical system and objective lens. A joint update of the speckle illumination during the reconstruction~\cite{yeh2019computational} could help to achieve more resolution gain with a low-NA objective for high-content imaging. Additionally, the calibrated speckle does not always match the actual illumination, making the reconstruction noisy as in Sec.~\ref{sec:experiment_results}. This mismatch can be prominent when the speckle contains features beyond the diffraction-limit of the system. Any changes between the calibration and the actual acquisition (e.g., phase shift on the illumination path) unaccounted for may also contribute to the speckle mismatch.

\section{Conclusion and Future Work}
\label{sec:conclusion}

We demonstrate a super-resolution method for dynamic scenes, called Speckle Flow SIM, which illuminates the sample with speckle-structured illumination to observe high frequency information beyond the diffraction limit. Speckle Flow SIM does not change its illumination but relies on the dynamics of the scene to acquire diversified measurements. This enables a simple, inexpensive experimental setup. Without loss of generality, we model the spatio-temporal relationship of the dynamic scene using the neural space-time model with coordinate-based multi-layer perceptrons, which jointly recover the motion dynamics and the super-resolved scene from a temporal sequence of raw images. We validate the Speckle Flow SIM in simulation and experiment. We show that Speckle Flow SIM can reconstruct a scene with deformable motion. We also experimentally demonstrate $1.88\times$ of the diffraction-limited resolution for a dynamic scene.

Future work may extend Speckle Flow SIM into the fluorescence channel where super-resolution microscopy methods are more commonly used by biologists and neuroscienists. A high-NA system is also needed for $\sim$100 nm spatial resolution. Another future direction is to enable the dynamic imaging (i.e., improve the temporal resolution) for other multi-shot computational imaging systems using the neural space-time model. By plugging in different forward models, the space-time model may jointly estimate the scene and the motion dynamics when the motion is relatively smooth. Besides, the neural space-time model itself can be further optimized for better computational efficiency and less network-induced reconstruction artifacts. Even though our space-time model has a great flexibility in its model selection, it is not exploited in this study, and our model was determined ad-hoc from limited tries. A systematic search of the neural network architecture~\cite{elsken2019neural} and hyperparameters~\cite{feurer2019hyperparameter} may be beneficial.

% % use section* for acknowledgment
% \ifCLASSOPTIONcompsoc
%   % The Computer Society usually uses the plural form
%   \section*{Acknowledgments}
% \else
%   % regular IEEE prefers the singular form
% \fi


\ifCLASSOPTIONcaptionsoff
  \newpage
\fi


\bibliographystyle{IEEEtran}
\bibliography{IEEEabrv,main}

\begin{thebibliography}{10}
\providecommand{\url}[1]{#1}
\csname url@samestyle\endcsname
\providecommand{\newblock}{\relax}
\providecommand{\bibinfo}[2]{#2}
\providecommand{\BIBentrySTDinterwordspacing}{\spaceskip=0pt\relax}
\providecommand{\BIBentryALTinterwordstretchfactor}{4}
\providecommand{\BIBentryALTinterwordspacing}{\spaceskip=\fontdimen2\font plus
\BIBentryALTinterwordstretchfactor\fontdimen3\font minus
  \fontdimen4\font\relax}
\providecommand{\BIBforeignlanguage}[2]{{%
\expandafter\ifx\csname l@#1\endcsname\relax
\typeout{** WARNING: IEEEtran.bst: No hyphenation pattern has been}%
\typeout{** loaded for the language `#1'. Using the pattern for}%
\typeout{** the default language instead.}%
\else
\language=\csname l@#1\endcsname
\fi
#2}}
\providecommand{\BIBdecl}{\relax}
\BIBdecl

\bibitem{mockl2014super}
L.~M{\"o}ckl, D.~C. Lamb, and C.~Br{\"a}uchle, ``Super-resolved fluorescence
  microscopy: nobel prize in chemistry 2014 for eric betzig, stefan hell, and
  william e. moerner,'' \emph{Angewandte Chemie International Edition},
  vol.~53, no.~51, pp. 13\,972--13\,977, 2014.

\bibitem{wu2018faster}
Y.~Wu and H.~Shroff, ``Faster, sharper, and deeper: structured illumination
  microscopy for biological imaging,'' \emph{Nature methods}, vol.~15, no.~12,
  pp. 1011--1019, 2018.

\bibitem{gustafsson2000surpassing}
M.~G. Gustafsson, ``Surpassing the lateral resolution limit by a factor of two
  using structured illumination microscopy,'' \emph{Journal of microscopy},
  vol. 198, no.~2, pp. 82--87, 2000.

\bibitem{hell1994breaking}
S.~W. Hell and J.~Wichmann, ``Breaking the diffraction resolution limit by
  stimulated emission: stimulated-emission-depletion fluorescence microscopy,''
  \emph{Optics letters}, vol.~19, no.~11, pp. 780--782, 1994.

\bibitem{betzig2006imaging}
E.~Betzig, G.~H. Patterson, R.~Sougrat, O.~W. Lindwasser, S.~Olenych, J.~S.
  Bonifacino, M.~W. Davidson, J.~Lippincott-Schwartz, and H.~F. Hess, ``Imaging
  intracellular fluorescent proteins at nanometer resolution,'' \emph{science},
  vol. 313, no. 5793, pp. 1642--1645, 2006.

\bibitem{godin2014super}
A.~G. Godin, B.~Lounis, and L.~Cognet, ``Super-resolution microscopy approaches
  for live cell imaging,'' \emph{Biophysical journal}, vol. 107, no.~8, pp.
  1777--1784, 2014.

\bibitem{chowdhury2013structured}
S.~Chowdhury and J.~Izatt, ``Structured illumination quantitative phase
  microscopy for enhanced resolution amplitude and phase imaging,''
  \emph{Biomedical optics express}, vol.~4, no.~10, pp. 1795--1805, 2013.

\bibitem{mudry2012structured}
E.~Mudry, K.~Belkebir, J.~Girard, J.~Savatier, E.~Le~Moal, C.~Nicoletti,
  M.~Allain, and A.~Sentenac, ``Structured illumination microscopy using
  unknown speckle patterns,'' \emph{Nature Photonics}, vol.~6, no.~5, pp.
  312--315, 2012.

\bibitem{wicker2013non}
K.~Wicker, ``Non-iterative determination of pattern phase in structured
  illumination microscopy using auto-correlations in fourier space,''
  \emph{Optics express}, vol.~21, no.~21, pp. 24\,692--24\,701, 2013.

\bibitem{lal2016structured}
A.~Lal, C.~Shan, and P.~Xi, ``Structured illumination microscopy image
  reconstruction algorithm,'' \emph{IEEE Journal of Selected Topics in Quantum
  Electronics}, vol.~22, no.~4, pp. 50--63, 2016.

\bibitem{yeh2019computational}
L.-H. Yeh, S.~Chowdhury, and L.~Waller, ``Computational structured illumination
  for high-content fluorescence and phase microscopy,'' \emph{Biomedical optics
  express}, vol.~10, no.~4, pp. 1978--1998, 2019.

\bibitem{yeh2019speckle}
L.-H. Yeh, S.~Chowdhury, N.~A. Repina, and L.~Waller, ``Speckle-structured
  illumination for 3d phase and fluorescence computational microscopy,''
  \emph{Biomedical optics express}, vol.~10, no.~7, pp. 3635--3653, 2019.

\bibitem{forster2016motion}
R.~F{\"o}rster, K.~Wicker, W.~M{\"u}ller, A.~Jost, and R.~Heintzmann, ``Motion
  artefact detection in structured illumination microscopy for live cell
  imaging,'' \emph{Optics Express}, vol.~24, no.~19, pp. 22\,121--22\,134,
  2016.

\bibitem{lucas2018using}
A.~Lucas, M.~Iliadis, R.~Molina, and A.~K. Katsaggelos, ``Using deep neural
  networks for inverse problems in imaging: beyond analytical methods,''
  \emph{IEEE Signal Processing Magazine}, vol.~35, no.~1, pp. 20--36, 2018.

\bibitem{stanley2007compositional}
K.~O. Stanley, ``Compositional pattern producing networks: A novel abstraction
  of development,'' \emph{Genetic programming and evolvable machines}, vol.~8,
  no.~2, pp. 131--162, 2007.

\bibitem{mildenhall2020nerf}
B.~Mildenhall, P.~P. Srinivasan, M.~Tancik, J.~T. Barron, R.~Ramamoorthi, and
  R.~Ng, ``Nerf: Representing scenes as neural radiance fields for view
  synthesis,'' in \emph{European conference on computer vision}.\hskip 1em plus
  0.5em minus 0.4em\relax Springer, 2020, pp. 405--421.

\bibitem{sitzmann2019scene}
V.~Sitzmann, M.~Zollh{\"o}fer, and G.~Wetzstein, ``Scene representation
  networks: Continuous 3d-structure-aware neural scene representations,''
  \emph{Advances in Neural Information Processing Systems}, vol.~32, 2019.

\bibitem{park2019deepsdf}
J.~J. Park, P.~Florence, J.~Straub, R.~Newcombe, and S.~Lovegrove, ``Deepsdf:
  Learning continuous signed distance functions for shape representation,'' in
  \emph{Proceedings of the IEEE/CVF Conference on Computer Vision and Pattern
  Recognition}, 2019, pp. 165--174.

\bibitem{park2021nerfies}
K.~Park, U.~Sinha, J.~T. Barron, S.~Bouaziz, D.~B. Goldman, S.~M. Seitz, and
  R.~Martin-Brualla, ``Nerfies: Deformable neural radiance fields,'' in
  \emph{Proceedings of the IEEE/CVF International Conference on Computer
  Vision}, 2021, pp. 5865--5874.

\bibitem{pumarola2021d}
A.~Pumarola, E.~Corona, G.~Pons-Moll, and F.~Moreno-Noguer, ``D-nerf: Neural
  radiance fields for dynamic scenes,'' in \emph{Proceedings of the IEEE/CVF
  Conference on Computer Vision and Pattern Recognition}, 2021, pp.
  10\,318--10\,327.

\bibitem{gustafsson2005nonlinear}
M.~G. Gustafsson, ``Nonlinear structured-illumination microscopy: wide-field
  fluorescence imaging with theoretically unlimited resolution,''
  \emph{Proceedings of the National Academy of Sciences}, vol. 102, no.~37, pp.
  13\,081--13\,086, 2005.

\bibitem{mangeat2021super}
T.~Mangeat, S.~Labouesse, M.~Allain, A.~Negash, E.~Martin, A.~Gu{\'e}nol{\'e},
  R.~Poincloux, C.~Estibal, A.~Bouissou, S.~Cantaloube \emph{et~al.},
  ``Super-resolved live-cell imaging using random illumination microscopy,''
  \emph{Cell Reports Methods}, vol.~1, no.~1, p. 100009, 2021.

\bibitem{yeh2017structured}
L.-H. Yeh, L.~Tian, and L.~Waller, ``Structured illumination microscopy with
  unknown patterns and a statistical prior,'' \emph{Biomedical optics express},
  vol.~8, no.~2, pp. 695--711, 2017.

\bibitem{kner2009super}
P.~Kner, B.~B. Chhun, E.~R. Griffis, L.~Winoto, and M.~G. Gustafsson,
  ``Super-resolution video microscopy of live cells by structured
  illumination,'' \emph{Nature methods}, vol.~6, no.~5, pp. 339--342, 2009.

\bibitem{lu2015fastsim}
H.-W. Lu-Walther, M.~Kielhorn, R.~F{\"o}rster, A.~Jost, K.~Wicker, and
  R.~Heintzmann, ``fastsim: a practical implementation of fast structured
  illumination microscopy,'' \emph{Methods and Applications in Fluorescence},
  vol.~3, no.~1, p. 014001, 2015.

\bibitem{shroff2010lateral}
S.~A. Shroff, J.~R. Fienup, and D.~R. Williams, ``Lateral superresolution using
  a posteriori phase shift estimation for a moving object: experimental
  results,'' \emph{JOSA A}, vol.~27, no.~8, pp. 1770--1782, 2010.

\bibitem{turcotte2019dynamic}
R.~Turcotte, Y.~Liang, M.~Tanimoto, Q.~Zhang, Z.~Li, M.~Koyama, E.~Betzig, and
  N.~Ji, ``Dynamic super-resolution structured illumination imaging in the
  living brain,'' \emph{Proceedings of the National Academy of Sciences}, vol.
  116, no.~19, pp. 9586--9591, 2019.

\bibitem{hasinoff2016burst}
S.~W. Hasinoff, D.~Sharlet, R.~Geiss, A.~Adams, J.~T. Barron, F.~Kainz,
  J.~Chen, and M.~Levoy, ``Burst photography for high dynamic range and
  low-light imaging on mobile cameras,'' \emph{ACM Transactions on Graphics
  (ToG)}, vol.~35, no.~6, pp. 1--12, 2016.

\bibitem{kellman2018motion}
M.~Kellman, M.~Chen, Z.~F. Phillips, M.~Lustig, and L.~Waller,
  ``Motion-resolved quantitative phase imaging,'' \emph{Biomedical Optics
  Express}, vol.~9, no.~11, pp. 5456--5466, 2018.

\bibitem{pnevmatikakis2016simultaneous}
E.~A. Pnevmatikakis, D.~Soudry, Y.~Gao, T.~A. Machado, J.~Merel, D.~Pfau,
  T.~Reardon, Y.~Mu, C.~Lacefield, W.~Yang \emph{et~al.}, ``Simultaneous
  denoising, deconvolution, and demixing of calcium imaging data,''
  \emph{Neuron}, vol.~89, no.~2, pp. 285--299, 2016.

\bibitem{sitzmann2020implicit}
V.~Sitzmann, J.~Martel, A.~Bergman, D.~Lindell, and G.~Wetzstein, ``Implicit
  neural representations with periodic activation functions,'' \emph{Advances
  in Neural Information Processing Systems}, vol.~33, pp. 7462--7473, 2020.

\bibitem{martel2021acorn}
J.~N. Martel, D.~B. Lindell, C.~Z. Lin, E.~R. Chan, M.~Monteiro, and
  G.~Wetzstein, ``Acorn: Adaptive coordinate networks for neural scene
  representation,'' \emph{arXiv preprint arXiv:2105.02788}, 2021.

\bibitem{raissi2019physics}
M.~Raissi, P.~Perdikaris, and G.~E. Karniadakis, ``Physics-informed neural
  networks: A deep learning framework for solving forward and inverse problems
  involving nonlinear partial differential equations,'' \emph{Journal of
  Computational physics}, vol. 378, pp. 686--707, 2019.

\bibitem{tancik2020fourier}
M.~Tancik, P.~Srinivasan, B.~Mildenhall, S.~Fridovich-Keil, N.~Raghavan,
  U.~Singhal, R.~Ramamoorthi, J.~Barron, and R.~Ng, ``Fourier features let
  networks learn high frequency functions in low dimensional domains,''
  \emph{Advances in Neural Information Processing Systems}, vol.~33, pp.
  7537--7547, 2020.

\bibitem{sun2021coil}
Y.~Sun, J.~Liu, M.~Xie, B.~Wohlberg, and U.~S. Kamilov, ``Coil:
  Coordinate-based internal learning for tomographic imaging,'' \emph{IEEE
  Transactions on Computational Imaging}, vol.~7, pp. 1400--1412, 2021.

\bibitem{liu2021zero}
R.~Liu, Y.~Sun, J.~Zhu, L.~Tian, and U.~Kamilov, ``Zero-shot learning of
  continuous 3d refractive index maps from discrete intensity-only
  measurements,'' \emph{arXiv preprint arXiv:2112.00002}, 2021.

\bibitem{hornik1989multilayer}
K.~Hornik, M.~Stinchcombe, and H.~White, ``Multilayer feedforward networks are
  universal approximators,'' \emph{Neural networks}, vol.~2, no.~5, pp.
  359--366, 1989.

\bibitem{bradbury2018jax}
J.~Bradbury, R.~Frostig, P.~Hawkins, M.~J. Johnson, C.~Leary, D.~Maclaurin, and
  S.~Wanderman-Milne, ``Jax: composable transformations of python+ numpy
  programs,'' \emph{Version 0.1}, vol.~55, 2018.

\bibitem{allen2001phase}
L.~Allen and M.~Oxley, ``Phase retrieval from series of images obtained by
  defocus variation,'' \emph{Optics communications}, vol. 199, no. 1-4, pp.
  65--75, 2001.

\bibitem{goodman2005introduction}
J.~W. Goodman, ``Introduction to fourier optics. 3rd,'' \emph{Roberts and
  Company Publishers}, 2005.

\bibitem{cragg2000lateral}
G.~E. Cragg and P.~T. So, ``Lateral resolution enhancement with standing
  evanescent waves,'' \emph{Optics letters}, vol.~25, no.~1, pp. 46--48, 2000.

\bibitem{yu2021plenoxels}
A.~Yu, S.~Fridovich-Keil, M.~Tancik, Q.~Chen, B.~Recht, and A.~Kanazawa,
  ``Plenoxels: Radiance fields without neural networks,'' \emph{arXiv preprint
  arXiv:2112.05131}, 2021.

\bibitem{mueller2022instant}
T.~M\"uller, A.~Evans, C.~Schied, and A.~Keller, ``Instant neural graphics
  primitives with a multiresolution hash encoding,'' \emph{arXiv:2201.05989},
  Jan. 2022.

\bibitem{elsken2019neural}
T.~Elsken, J.~H. Metzen, and F.~Hutter, ``Neural architecture search: A
  survey,'' \emph{The Journal of Machine Learning Research}, vol.~20, no.~1,
  pp. 1997--2017, 2019.

\bibitem{feurer2019hyperparameter}
M.~Feurer and F.~Hutter, ``Hyperparameter optimization,'' in \emph{Automated
  machine learning}.\hskip 1em plus 0.5em minus 0.4em\relax Springer, Cham,
  2019, pp. 3--33.

\end{thebibliography}

\ifpeerreview \else
%%%% For the camera ready version, please fill out this
%%%% biography. Your camera ready should be within a 12 page limit
%%%% including acknowledgments, references and biography.

% \includegraphics command within an optional argument. (You could
% create your own custom macro containing the \includegraphics command
% to make things simpler here.)
% \begin{IEEEbiography}[{\includegraphics[width=1in,height=1.25in,clip,keepaspectratio]{mshell}}]{Michael Shell}

\begin{IEEEbiography}[{\includegraphics[width=1in,height=1.25in,clip,keepaspectratio]{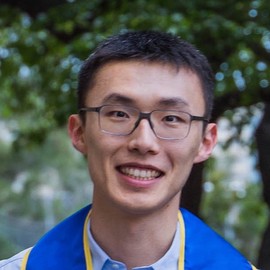}}]{Ruiming Cao} is a Ph.D. candidate in Bioengineering at University of California, Berkeley, advised by Prof. Laura Waller. He received B.S. in Computer Science and Applied Mathematics and M.S. in Computer Science from University of California, Los Angeles. He was an research intern at Google Research in 2021 and an optical scientist intern in Meta Reality Labs Research in 2022. His research is on the intersection of computational imaging, optics, and biomedical imaging.
\end{IEEEbiography}

\begin{IEEEbiography}[{\includegraphics[width=1in,height=1.25in,clip,keepaspectratio]{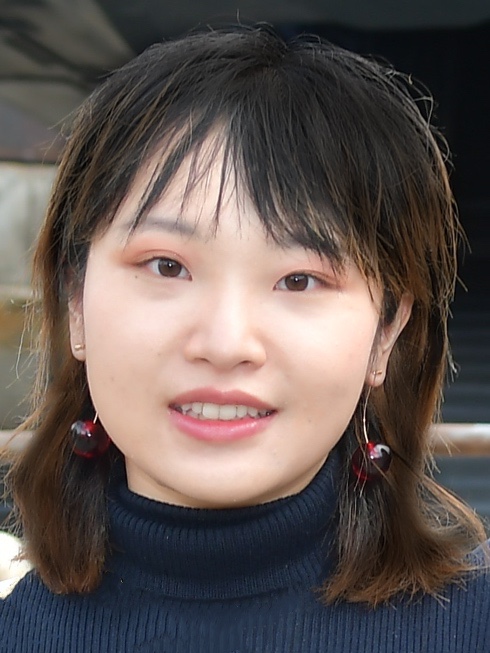}}]{Fanglin Linda Liu}
received the B.S. degree from Tsinghua University, Beijing, China, in 2016, and the Ph.D. degree in electrical engineering and computer science from the University of California, Berkeley, Berkeley, CA, USA, in 2022. She was an optical scientist intern at Microsoft Research, Cambridge, UK, in 2021, and joined Google X, the moonshot factory as a hardware engineer in 2022. Her research focuses on the design of computational imaging systems for optical microscopy and 3D imaging.
\end{IEEEbiography}

\begin{IEEEbiography}[{\includegraphics[width=1in,height=1.25in,clip,keepaspectratio]{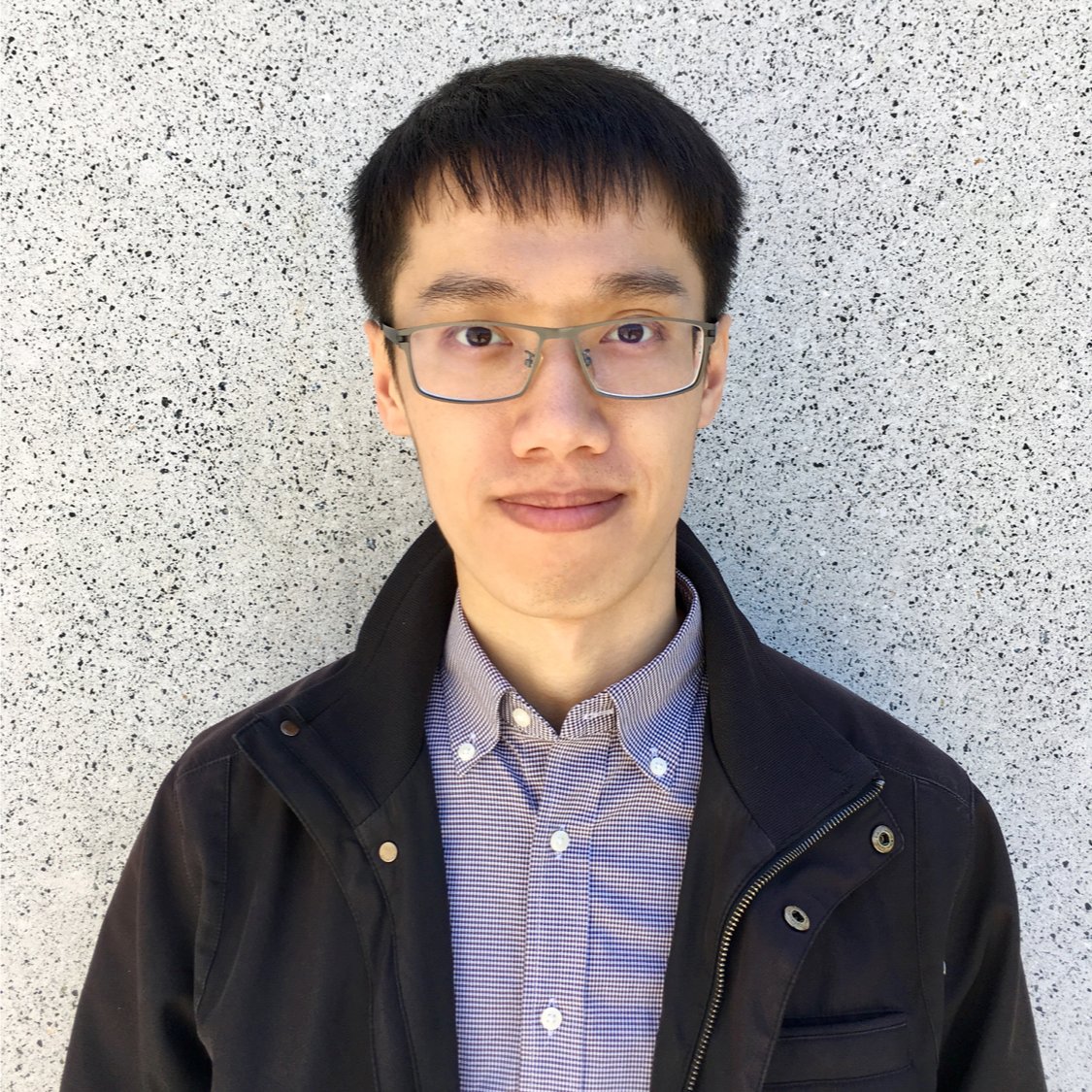}}]{Li-Hao Yeh}
received a B.S. degree in electrical engineering from National Taiwan University, Taipei, Taiwan, in 2013, the M.S. and Ph.D. degrees in EECS from the University of California, Berkeley, in 2016, and 2019, respectively. 

He is currently a Senior Engineer at ASML. Prior to joining ASML, he was an R\&D Engineer in Chan Zuckerburg Biohub from 2019 to 2022. His research interest focuses on computational imaging in the applications of label-free microscopy and lithography.
\end{IEEEbiography}

\begin{IEEEbiography}[{\includegraphics[width=1in,height=1.25in,clip,keepaspectratio]{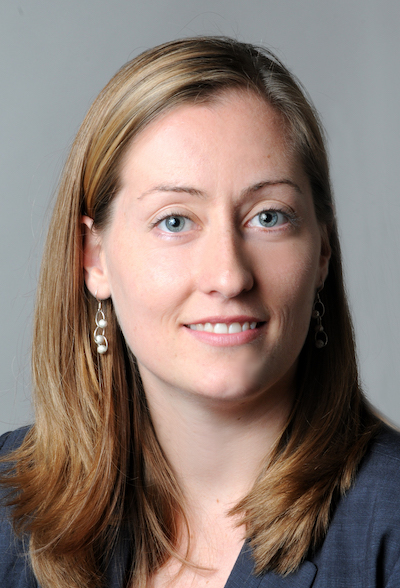}}]{Laura Waller}
is a Professor of Electrical Engineering and Computer Sciences at UC Berkeley. She received B.S., M.Eng. and Ph.D. degrees from the Massachusetts Institute of Technology and was a Postdoc/Lecturer of Physics at Princeton University. She is a Packard Fellow, Moore Foundation Data-driven Investigator, Bakar Fellow, OSA Fellow, AIMBE Fellow and Chan-Zuckerberg Biohub Investigator. She received the Carol D.Soc Distinguished Graduate Mentoring Award, OSA Adolph Lomb Medal, Ted Van Duzer Endowed Professorship, NSF CAREER Award and SPIE Early Career Achievement Award. 
\end{IEEEbiography}

%\vfill

\fi\end{document}